\documentclass[]{aa}
\usepackage{graphicx}
\usepackage{txfonts}
\usepackage{epstopdf}

\begin{document}

\title{Generalised model-independent characterisation of strong gravitational lenses II: Transformation matrix between multiple images}
\titlerunning{Model-independent characterisation of strong gravitational lenses II: Transformation matrix}
\author{J. Wagner\inst{1} \and N. Tessore\inst{2}}
\institute{Universit\"at Heidelberg, Zentrum f\"ur Astronomie, Institut f\"ur Theoretische Astrophysik, Philosophenweg 12, 69120 Heidelberg, Germany, Heidelberg Institute for Theoretical Studies, 69118 Heidelberg, Germany \\
\email{j.wagner@uni-heidelberg.de}
\and
Jodrell Bank Centre for Astrophysics, School of Physics and Astronomy, University of Manchester, Alan Turing Building, Oxford Road, Manchester M13 9PL, UK \\ \email{nicolas.tessore@manchester.co.uk}}
\date{Received XX; accepted XX}

\abstract{We determine the transformation matrix that maps multiple images with identifyable resolved features onto one another and that is based on a Taylor-expanded lensing potential in the vicinity of a point on the critical curve within our model-independent lens characterisation approach. From the transformation matrix, the same information about the properties of the critical curve at fold and cusp points can be derived as we previously found when using the quadrupole moment of the individual images as observables. In addition, we read off the relative parities between the images, so that the parity of all images is determined, when one is known. We compare all retrievable ratios of potential derivatives to the actual ones and to the ones obtained by using the quadrupole moment as observable for two and three image configurations generated by a galaxy-cluster scale singular isothermal ellipse. We conclude that using the quadrupole moments as observables, the properties of the critical curve at the cusp points are retrieved to a higher accuracy, at the fold points to a lower accuracy, and the ratios of second order potential derivatives to comparable accuracy.
We also show that the approach using ratios of convergences and reduced shear components is equivalent to ours in the vicinity of the critical curve but yields more accurate results and is more robust because it does not require a special coordinate system as the approach using potential derivatives does.
The transformation matrix is determined by mapping manually assigned reference points in the multiple images onto one another. If the assignment of the reference points is subject to measurement uncertainties under the influence of noise, we find that the confidence intervals of the lens parameters can be as large as the values themselves, when the uncertainties are larger than one pixel. In addition, observed multiple images with resolved features are more extended than unresolved ones, so that higher order moments should be taken into account to improve the reconstruction precision and accuracy.}

\maketitle

%%%%%%%%%%%%%%%%%
\section{Introduction}
\label{sec:introduction}

High-redshift galaxies can be magnified by strong gravitational lensing such that their multiple images show resolved features, as for instance observed by \cite{bib:Colley, bib:Donnarumma, bib:Sharon}. In this way, the properties and evolution of these faint and otherwise hardly observable galaxies can be studied. When resolved features within several of the multiple images can be identified, \cite{bib:Tessore} established a general expression to link the transformations between any configuration of more than two multiple images with properties of the lens mapping. Information about the ratios of convergences and the reduced shear at the position of the images can then be obtained without assuming any lens model. In this work, we investigate which model-independent information about the gravitational lens can be retrieved from the transformation between multiple images with resolved features in terms of a Taylor-expanded lens potential, as first introduced in \cite{bib:Wagner}, in fold and cusp configurations close to the critical curve.

In Section~\ref{sec:transformation} we revise the approach based on the Taylor-expanded lensing potential as developed in \cite{bib:Wagner, bib:Wagner2}, and the approach established in \cite{bib:Tessore} in a unified notation. We show that both approaches are equivalent for fold and cusp configurations close to the critical curve. Subsequently we compare and combine them to extend the model-independent knowledge retrievable from multiple image configurations. After the theoretical derivations, we discuss the accuracy and precision achievable and, in Section~\ref{sec:implementation}, briefly introduce the algorithmic implementation to obtain the lens parameters, i.e.\ the ratios of convergences, reduced shear components or ratios of potential derivatives, from the linear transformation between resolved features in multiple images. It will be further detailed in \cite{bib:Tessore2}.

\begin{figure*}[ht!]
\centering
  \includegraphics[width=0.8\textwidth]{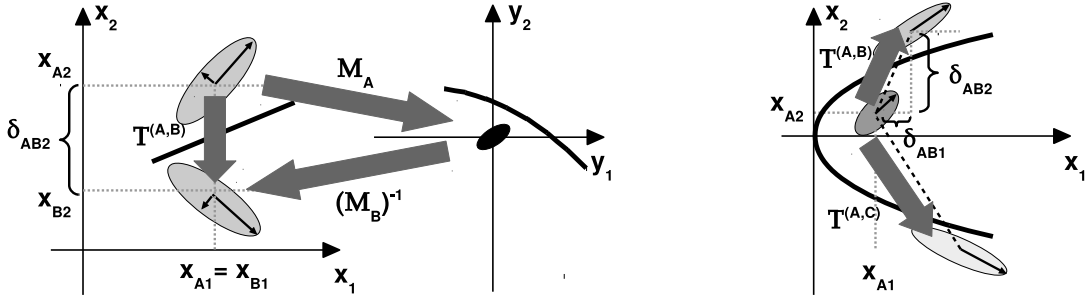} % problems with 1.0!
    \caption{Transformation $T^{(A,B)} = M_B^{-1} M_A$ between image $A$ and $B$ at a fold singular point (left) assembled from projecting $A$ back to the source using $M_A$ and projecting the source onto $B$ using $M_B^{-1}$ (left, centre). Transformations between three images at a cusp singular point, projecting the central image $A$ onto $B$ using $T^{(A,B)} = M_B^{-1} M_A$ and projecting $A$ onto $C$ using $T^{(A,C)} = M_C^{-1} M_A$ (right). (Not shown is the transformation between $B$ and $C$, e.g. $T^{(B,C)} = M_C^{-1} M_B$.)} 
 \label{fig:transformation}
\end{figure*}

Section~\ref{sec:examples} then shows an applicational example: a simulated set of four multiple images in a singular isothermal elliptical (SIE) lens model. By means of this simulation we analyse how accurately the lens parameters can be determined for the different choices of variables of \cite{bib:Tessore} and \cite{bib:Wagner2} in more detail and compare the results with the results that are gained when using the quadrupole moment of the individual images as observable instead of the transformation matrix. We furthermore investigate the influence of the size of the multiple images and the influence of detection noise on the accuracy of the lens parameter reconstruction. 

Section~\ref{sec:conclusion} summarises the results and gives an outlook to the observational cases that can be analysed with this approach.

%%%%%%%%%%%%%%%%%%
\section{Transformation matrix for a Taylor-expanded lensing potential}
\label{sec:transformation}

\subsection{Definitions and notations}
\label{sec:definitions}

Let $\phi(\boldsymbol{x},\boldsymbol{y})$ be the gravitational lensing potential that defines the lens mapping between $\boldsymbol{x} \in \mathbb{R}^2$ in the image plane and $\boldsymbol{y} \in \mathbb{R}^2$ in the source plane by $\nabla_{\boldsymbol{x}} \phi(\boldsymbol{x},\boldsymbol{y}) = 0$. The critical curves are all points $\boldsymbol{x}_{0}$ for which the lens mapping becomes singular. Mapping them into the source plane, we obtain the caustic points $\boldsymbol{y}_{0}$.
As already introduced in \cite{bib:SEF, bib:Wagner, bib:Wagner2}, the most convenient coordinate system to characterise the lens mapping in the vicinity of the critical curve using a Taylor-expansion of $\phi$ around a point $\boldsymbol{x}_{0}$ on the critical curve is given by the conditions 
\begin{equation}
\boldsymbol{x}_{0}= (0,0) \;, \quad  \boldsymbol{y}_{0} = (0,0)\;, \quad  \phi_1^{(0)} = \phi_2^{(0)} = \phi_{12}^{(0)} = \phi_{22}^{(0)} = 0 \;.
\label{eq:coordinate_system}
\end{equation}
A subscript $i$ of $\phi$ denotes the partial derivative in the direction of $x_i$, $i=1,2$, and the superscript $(0)$ indicates that this variable is evaluated at the singular point. Analogously, superscripts of capital Latin letters, $(A), (B), ...$, denote that the variable is considered at the centre of light of the respective image. Furthermore, we define the relative distance between the centres of light of image $I$ and $J$, $\boldsymbol{x}_{I}$ and $\boldsymbol{x}_J$, as $\boldsymbol{\delta}_{IJ} = \left( \delta_{IJ1}, \delta_{IJ2} \right) = \left( x_{I1} - x_{J1}, x_{I2} - x_{J2} \right)$.
%\begin{equation}
%\boldsymbol{\delta}_{IJ} = \left( \delta_{IJ1}, \delta_{IJ2} \right) = \left( x_{I1} - x_{J1}, x_{I2} - x_{J2} \right)\;.
%\end{equation}

Given two images $A$ and $B$ and their distortion matrices, often also called magnification matrices, $M_A(\boldsymbol{x})$ and $M_B(\boldsymbol{x})$, we define the transformation matrix, also called relative magnification matrix, $T$ that maps image $A$ onto image $B$ by
\begin{equation}
T^{(A,B)} (\boldsymbol{x}) = \left( M_B^{-1} \circ M_A \right) (\boldsymbol{x}) \;,
\label{eq:transformation_matrix}
\end{equation}
as also stated in \cite{bib:Gorenstein, bib:Narayan}. The transformation matrix between the two images can be interpreted as a linear approximation of the map that first projects points of image $A$, $\boldsymbol{x}_{\{A\}}$, into the source plane by $\boldsymbol{y}(\boldsymbol{x}_{\{A\}}) \approx M_A (\boldsymbol{x}_{\{A\}})$ and subsequently maps the source onto the points in image $B$ by $\boldsymbol{x}_{\{B\}}(\boldsymbol{y}) \approx M_B^{-1}(\boldsymbol{y})$, as Figure~\ref{fig:transformation} visualises.

In the coordinate system given by Equation~\eqref{eq:coordinate_system} and using the Taylor-expanded lensing potential, the distortion matrices are given by
\begin{equation}
M_I = \left( \begin{matrix} \phi_{11}^{(I)} &  \phi_{12}^{(I)}  \\[1ex]  \phi_{12}^{(I)} &  \phi_{22}^{(I)} \end{matrix}\right) \;, \quad I = A, B, ...
\label{eq:distortion_matrix}
\end{equation}
with the matrix entries
\begin{align}
\phi_{11}^{(I)} &= \phi_{11}^{(0)} \;, \label{eq:rel_phis1} \\
\phi_{12}^{(I)} &= \phi_{122}^{(0)} x_{I2} \;, \label{eq:rel_phis2} \\
\phi_{22}^{(I)} &= \phi_{222}^{(0)} x_{I2} \;, \quad \text{(fold)} \label{eq:rel_phis3} \\
\phi_{22}^{(I)} &= \phi_{122}^{(0)} x_{I1} + \tfrac12 \phi_{2222}^{(0)} \left( x_{I2} \right)^2 \; \quad \text{(cusp)}\;.  \label{eq:rel_phis4}
\end{align}
in the vicinity of a fold or a cusp critical point $\boldsymbol{x}_{0}$ (see \cite{bib:Wagner2} for further details).

To be consistent with the notation given above, we reformulate the transformation equations, Equations (17)-(20) in \cite{bib:Tessore}, for $n$ multiple images of a gravitationally lensed source as
\begin{align}
T^{(A,I)}_{11} &=    \dfrac{1-\kappa^{(A)}}{1-\kappa^{(I)}}  \dfrac{\left(1-g_1^{(A)}\right)\left( 1 + g_1^{(I)} \right)-g_2^{(A)} g_2^{(I)}}{1-\left(g_1^{(I)}\right)^2 - \left(g_2^{(I)}\right)^2} \;, \quad I = B, C, ...\label{eq:T11} \\
T^{(A,I)}_{12} &=  \dfrac{1-\kappa^{(A)}}{1-\kappa^{(I)}}  \dfrac{\left(1+g_1^{(A)}\right) g_2^{(I)} - \left( 1 + g_1^{(I)} \right)g_2^{(A)}}{1-\left(g_1^{(I)}\right)^2 - \left(g_2^{(I)}\right)^2}  \;, \label{eq:T22}\\
T^{(A,I)}_{21} &=   \dfrac{1-\kappa^{(A)}}{1-\kappa^{(I)}}  \dfrac{\left(1-g_1^{(A)}\right) g_2^{(I)} - \left( 1 - g_1^{(I)} \right)g_2^{(A)}}{1-\left(g_1^{(I)}\right)^2 - \left(g_2^{(I)}\right)^2} \;, \label{eq:T21} \\
T^{(A,I)}_{22} &=   \dfrac{1-\kappa^{(A)}}{1-\kappa^{(I)}}  \dfrac{\left(1+g_1^{(A)}\right)\left( 1 - g_1^{(I)} \right)-g_2^{(A)} g_2^{(I)}}{1-\left(g_1^{(I)}\right)^2 - \left(g_2^{(I)}\right)^2} \;, \label{eq:T22}
\end{align}
in which the subscripts of $T$ $i,j = 1,2$ denote the entries of the transformation matrices between the reference image, called image $A$ without loss of generality, and the remaining $n-1$ multiple images $I$. As usual, $\kappa$ denotes the convergence and $\boldsymbol{g} = \boldsymbol{\gamma}/(1-\kappa)$ is the reduced shear, for which a subscript $i=1,2$ denotes its components, as defined in \cite{bib:Weak_lensing_review}.

%%%%%%%%%%%%%%%%%%
\subsection{Equivalence of both approaches}
\label{sec:equivalence_of_approaches}

The equivalent set of equations using the derivatives of the lensing potential as variables reads:
\begin{align}
T^{(A,I)}_{11} &= \dfrac{\phi_{11}^{(A)} }{\phi_{11}^{(I)}}  \dfrac{\left(\tilde{\phi}_{22}^{(I)} - \tilde{\phi}_{12}^{(A)} \tilde{\phi}_{12}^{(I)} \right)}{\tilde{\phi}_{22}^{(I)} - \left( \tilde{\phi}_{12}^{(I)} \right)^2} \;, \quad I = B, C, ...\label{eq:T11_phi} \\
T^{(A,I)}_{12} &= \dfrac{\phi_{11}^{(A)} }{\phi_{11}^{(I)}}  \dfrac{\left(\tilde{\phi}_{12}^{(A)} \tilde{\phi}_{22}^{(I)} - \tilde{\phi}_{22}^{(A)} \tilde{\phi}_{12}^{(I)} \right)}{\tilde{\phi}_{22}^{(I)} - \left( \tilde{\phi}_{12}^{(I)} \right)^2} \;, \label{eq:T22_phi}\\
T^{(A,I)}_{21} &=  \dfrac{\phi_{11}^{(A)} }{\phi_{11}^{(I)}}  \dfrac{\left(\tilde{\phi}_{12}^{(A)} - \tilde{\phi}_{12}^{(I)}\right)}{\tilde{\phi}_{22}^{(I)} - \left( \tilde{\phi}_{12}^{(I)} \right)^2} \;, \label{eq:T21_phi} \\
T^{(A,I)}_{22} &=  \dfrac{\phi_{11}^{(A)} }{\phi_{11}^{(I)}}  \dfrac{\left(\tilde{\phi}_{22}^{(A)} - \tilde{\phi}_{12}^{(A)} \tilde{\phi}_{12}^{(I)} \right)}{\tilde{\phi}_{22}^{(I)} - \left( \tilde{\phi}_{12}^{(I)} \right)^2} \;,\label{eq:T22_phi}
\end{align}
with the abbreviation
\begin{equation}
\tilde{\phi}_{ij}^{(I)} = \dfrac{\phi_{ij}^{(I)}}{\phi_{11}^{(I)}} \;, \quad (i,j) = (1,2), (2,2) \;, \quad I=A, B, C,... \;.
\end{equation}
Given at least three multiple images of one source with their transformation matrices and introducing the variable
\begin{equation}
f_\kappa^{(I)} \equiv \dfrac{1-\kappa^{(A)}}{1-\kappa^{(I)}}\;, \quad f_{\phi}^{(I)} \equiv \dfrac{\phi_{11}^{(A)}}{\phi_{11}^{(I)}} \;, \quad I=B, C, ...
\end{equation}
as in \cite{bib:Tessore}, both systems of equations can be solved, one for $f_\kappa^{(I)}, g_1^{(I)}, g_2^{(I)}$, the other for $f_\phi^{(I)}, \tilde{\phi}_{12}^{(I)}, \tilde{\phi}_{22}^{(I)}$. Both solutions are transformable into one another and unique. 

Yet, if all observable images are aligned and oriented orthogonal to the critical curves, both systems of equations are underdetermined. This is the case for all axisymmetric lensing configurations, e.g.\ generated by an Navarro-Frenk-White (NFW) profile. To demonstrate this degeneracy, consider the transformations between the three multiple images generated by an NFW profile. Their quadrupole moments are aligned with each other and extended orthogonally to the critical curve, implying that their transformation matrices are diagnoal. Hence, the systems of equations, Equations~\eqref{eq:T11} to \eqref{eq:T22} and Equations~\eqref{eq:T11_phi} to \eqref{eq:T22_phi} for the three images $A, B, C$, denoting the reference image as $A$, reduce to four equations (two equations of $T_{11}$ from the combination $(A,B)$ and $(A,C)$ and two equations of $T_{22}$ from the same combintions) to solve for five variables, $f_\kappa^{(B)}, f_\kappa^{(C)}, g_{1}^{(A)}, g_{1}^{(B)}, g_{1}^{(C)}$ or $f_\phi^{(B)}, f_\phi^{(C)}, \tilde{\phi}_{22}^{(A)}, \tilde{\phi}_{22}^{(B)}, \tilde{\phi}_{22}^{(C)}$. Hence, the system is underdetermined. Analogously, the transformation matrix cannot be employed for two images that straddle a fold and are oriented orthogonally to the critical curve, either. 

In the approach of \cite{bib:Tessore}, no special coordinate system is required, while the system established in Equations~\ref{eq:T11_phi} to \ref{eq:T22_phi} requires a coordinate system in which the images are extended along the $x_2$-axis. Furthermore, the direction of the semi-major axis of the image quadrupole moments remains unchanged by the coordinate transformation, i.e.\ the sign of the off-diagonal entries in the magnification matrix is kept. 
Appendix~\ref{app:phi_solution} shows the solution to the system of equations in Equations~\ref{eq:T11_phi} to \ref{eq:T22_phi} in these coordinates for three images in analogy to the one derived in \cite{bib:Tessore}. Yet, as detailed in Section~\ref{sec:implementation}, we pursue another way to solve for the lens parameters here.

Measuring the quadrupole moment of an image that need not have resolved features, the reduced shear and second order potential derivatives at its centre of light are determined by its axis ratio and orientation angle, as shown in \cite{bib:Wagner2}.

%%%%%%%%%%%%%%%%%%
\subsection{Transformations close to the critical curve}
\label{sec:derivation}

For two images $A$ and $B$ close to a point $\boldsymbol{x}_{0}$ on the critical curve, in the special coordinate system of Equation~\ref{eq:coordinate_system},
\begin{equation}
f_\kappa = \dfrac{1-\kappa^{(A)}}{1-\kappa^{(B)}} \approx 1\;, \quad | f_{\phi} | = \left| \dfrac{\phi_{11}^{(A)}}{\phi_{11}^{(B)}} \right| \approx \dfrac{\phi_{11}^{(0)}}{\phi_{11}^{(0)}} = 1
\label{eq:limit}
\end{equation}
holds, which eliminates one of the five variables in Equations~\eqref{eq:T11} to \eqref{eq:T22} or Equations~\eqref{eq:T11_phi} to \eqref{eq:T22_phi}, so that the system is exactly solved using the transformation matrix between the two images. If additional images are present, the validity of the limit in Equation~\eqref{eq:limit} can be tested by calculating $f_\phi$ from Equations~\eqref{eq:T11_phi} to \eqref{eq:T22_phi}. Small deviations from the limit imply that the two images are so close to the critical curve that the intrinsic source ellipticity and orientation to the caustic are negligible and observable image ellipticities lead to the results deduced in \cite{bib:Wagner2}. Analogously, small deviations from the limit imply that the following ratios of second order derivatives at the image positions can be determined when using the transformation matrix between the two images, as observable instead of their quadrupole moments
\begin{align}
\tilde{\phi}_{12}^{(A)} &= \dfrac{T_{22} (1 - T_{11}) + T_{12} T_{21}}{T_{21}} \;, \label{eq:phitilde_1} \\
\tilde{\phi}_{22}^{(A)} &= \dfrac{T_{22} (1-T_{11})(T_{22}-T_{11}) + T_{12} T_{21} (1-T_{11}+T_{22})}{T_{21}^2} \;, \label{eq:phitilde_2}
\end{align}
\begin{align}
\tilde{\phi}_{12}^{(B)} &= \dfrac{1-T_{11}}{T_{21}} \;, \label{eq:phitilde_3} \\
\tilde{\phi}_{22}^{(B)} &= -\dfrac{(T_{22}-T_{11}) (1 - T_{11}) + T_{12} T_{21}}{T_{21}^2} \;, \label{eq:phitilde_4}
\end{align}
in which we denote the image with positive parity as $A$ and assume $T_{21} \ne 0$. (For better readability, we drop the superscripts of $T$ when considering only one pair of images.) Inserting these relations into the lensing equations, we obtain
\begin{align}
\tilde{\phi}_{122}^{(0)} &\equiv \dfrac{\phi_{122}^{(0)}}{\phi_{11}^{(0)}} = \dfrac{2}{\delta_{AB2}} \tilde{\phi}^{(A)}_{12} = -\dfrac{2}{\delta_{AB2}} \tilde{\phi}^{(B)}_{12}\;, \label{eq:phi122_fold} \\
\tilde{\phi}_{222}^{(0)} &\equiv \dfrac{\phi_{222}^{(0)}}{\phi_{11}^{(0)}} = \dfrac{2}{\delta_{AB2}} \tilde{\phi}^{(A)}_{22} = \dfrac{2}{\delta_{AB2}} \tilde{\phi}^{(B)}_{22}
\label{eq:phi222_fold}
\end{align}
for the two images close to a fold.

For three images at a cusp $A$, $B$, $C$, with $A$ being the reference image closest to the cusp, we obtain
\begin{align}
\tilde{\phi}_{122}^{(0)} &\equiv \dfrac{\phi_{122}^{(0)}}{\phi_{11}^{(0)}} = \dfrac{1}{\delta_{AB2}} \left(\tilde{\phi}^{(A)}_{12} - \tilde{\phi}^{(B)}_{12} \right) \;, \\
\tilde{\phi}_{2222}^{(0)} &\equiv \dfrac{\phi_{2222}^{(0)}}{\phi_{11}^{(0)}} = -\dfrac{6}{\delta_{AB2}^2} \left( \dfrac{\delta_{AB1}}{\delta_{AB2}}\tilde{\phi}_{12}^{(B)} + \tilde{\phi}^{(A)}_{22} \right) \label{eq:phi2222_cusp} \;,
\end{align}
and analogous relations using images $A$ and $C$. Equation~\ref{eq:limit} has to be valid for all three images. Determining the $f_\phi^{(I)}$, $I=B,C$, shows the goodness of approximation of Equation~\ref{eq:limit}.

Details about the derivations of Equations~\eqref{eq:phi122_fold} to \eqref{eq:phi2222_cusp} are given in \cite{bib:Wagner2}, as the calculations employing the quadrupole moment are equivalent until the ratios of second order derivatives at the image positions are replaced by the observables. Hence, we obtain the same ratios of potential derivatives as results which determine the shape of the critical curve in the vicinity of the images as derived in \cite{bib:Wagner2}.

In addition, we simplify the reconstruction of the relative image position of $A$ established in \cite{bib:Wagner2}, using
\begin{align}
x_{A1} &= \left( \dfrac{\phi_{122}^{(0)}}{\phi_{11}^{(0)}} \right)^{-1} \left( \tilde{\phi}^{(A)}_{22} - \dfrac12 \dfrac{\phi_{2222}^{(0)}}{\phi_{11}^{(0)}} (x_{A2})^2 \right) \;, \\
x_{A2} &= \tilde{\phi}^{(A)}_{12} \left( \dfrac{\phi_{122}^{(0)}}{\phi_{11}^{(0)}} \right)^{-1} \;,
\end{align}
which is directly derived from Equations~\ref{eq:rel_phis2} and \ref{eq:rel_phis4}. Having determined $\boldsymbol{x}_A$, we know the position of the cusp critical point.

If $T_{21} = 0$, $\tilde{\phi}_{12}^{(A)} = \tilde{\phi}_{12}^{(B)}$, which implies that both are zero, considering Equation~\eqref{eq:rel_phis2}. From this follows that $T_{11} = 1$, $T_{12} = 0$ and $T_{22} =  \tilde{\phi}_{22}^{(A)}/  \tilde{\phi}_{22}^{(B)}$. As a consequence, we infer that the images are oriented orthogonal to the critical curve and that the latter equation is underdetermined, so that we cannot retrieve $\phi_{222}^{(0)}/\phi_{11}^{(0)}$.

Hence, if a transformation with $T_{21} = 0$ occurs that does not also have $T_{11} = 1$ and $T_{12} = 0$, the considered image configuration is inconsistent with our approach, which may hint at local asymmetries in the lensing potential, microlensing, dust extinction or to the fact that the two images may not originate from the same source, as already detailed in \cite{bib:Wagner2}. Given more than two images, i.e.\ an overdetermined system of equations, we can also use the surplus constraints to detect such anomalies. 

%%%%%%%%%%%%%%%%%%
\subsection{Relative parity information}
\label{sec:parity}

As can be easily derived from Equation~\ref{eq:transformation_matrix}
\begin{equation}
\det{\left( T \right)} = \dfrac{\mu_B}{\mu_A}
\end{equation}
holds for the magnification ratio $\mu_B/\mu_A$. Hence, the sign of the determinant of the transformation yields the relative parities between the images. Given the parity of one image, all parities can be fixed. For instance, assuming that the faint image of a source very close to the lens centre is a maximum in the lens mapping and thus of positive parity, it can be used to determine the parities of the remaining images. The five multiple images of a source at $z_\mathrm{s}=1.675$ in the galaxy cluster Cl0024+1654, \cite{bib:Colley} are an example for which the image parities can be determined in this way.

Deviations between the absolute values of $\det{(T)}$ and the observed magnification ratios may hint at microlensing or dust extinction.  

\begin{figure*}[h!]
\centering
  \includegraphics[width=\textwidth]{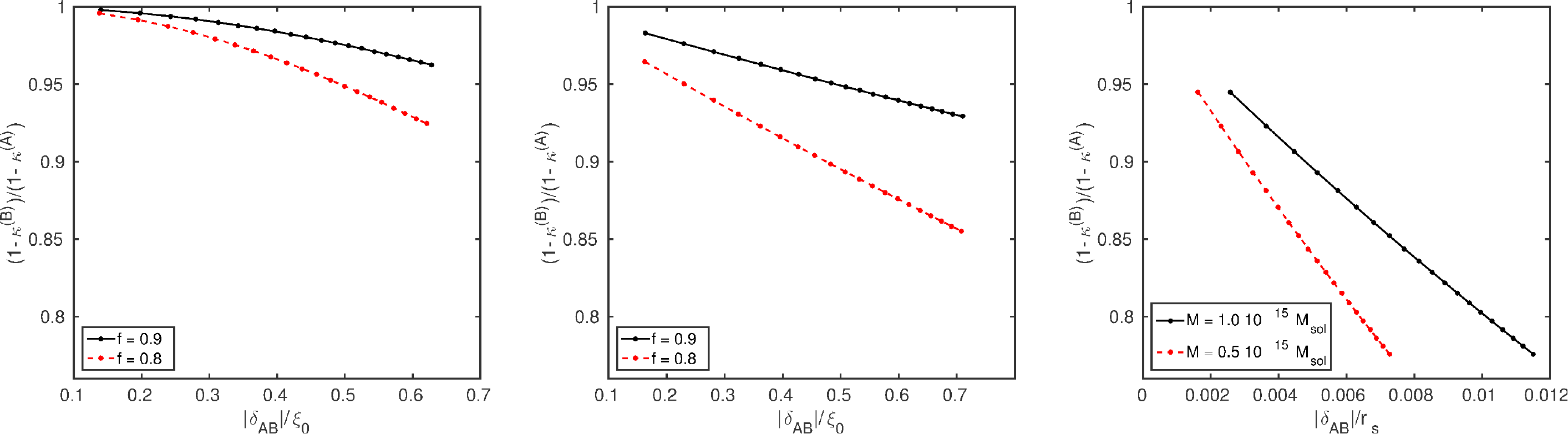}
    \caption{Accuracy of the approximation stated in Equation~\eqref{eq:limit} for the variables used in \cite{bib:Tessore} close to a symmetric cusp configuration of an SIE (left), close to a fold configuration of an SIE (centre) and close to the fold configuration of an NFW profile (right). Distances between the images $|\boldsymbol{\delta}_{AB}|$ are calculated with respect to the scale radius of the respective model. Black solid lines denote the accuracy for an SIE with $f=0.9$ or an NFW with $M = 1.0 \cdot 10^{15} M_\odot$, red dashed lines for an SIE with $f=0.8$ or an NFW with $M = 0.5 \cdot 10^{15} M_\odot$. (Notation for the models as defined in \cite{bib:Kormann} and \cite{bib:Bartelmann}.)} 
    %$\xi_0 = 4\pi \left( \tfrac{v}{c}\right)^2 \dfrac{D_\mathrm{d} D_\mathrm{ds}}{D_\mathrm{s}}$, $r_s = r_{200}/c_\mathrm{NFW}$.}
    \label{fig:accuracy_comparison}
\end{figure*}

\begin{figure*}[h!]
\centering
  \includegraphics[width=\textwidth]{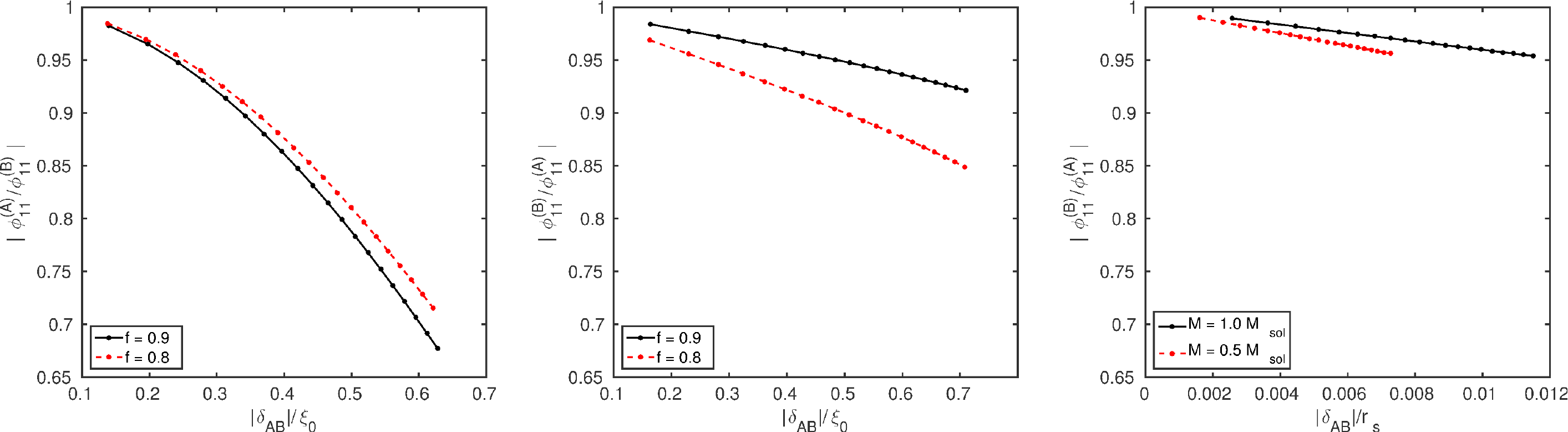}
    \caption{Accuracy of the approximation stated in Equation~\eqref{eq:limit} for the variables using in \cite{bib:Wagner2} close to a symmetric cusp configuration of an SIE (left, notice that $A$ and $B$ are interchanged), close to a fold configuration of an SIE (centre) and close to the fold configuration of an NFW profile (right). Distances between the images $|\boldsymbol{\delta}_{AB}|$ are calculated with respect to the scale radius of the respective model. Black solid lines denote the accuracy for an SIE with $f=0.9$ or an NFW with $M = 1.0 \cdot 10^{15} M_\odot$, red dashed lines for an SIE with $f=0.8$ or an NFW with $M = 0.5 \cdot 10^{15} M_\odot$. (Notation for the models as defined in \cite{bib:Kormann} and \cite{bib:Bartelmann}.)} 
    %$\xi_0 = 4\pi \left( \tfrac{v}{c}\right)^2 \dfrac{D_\mathrm{d} D_\mathrm{ds}}{D_\mathrm{s}}$, $r_s = r_{200}/c_\mathrm{NFW}$.}
    \label{fig:accuracy_comparison_phi}
\end{figure*}

%%%%%%%%%%%%%%%%%%
\subsection{Accuracy and precision}
\label{sec:errors}

The results of \cite{bib:Tessore} are derived for the idealised case of a transformation between images at infinite resolution in the absence of noise. In realistic cases, assumptions have to be made to deal with finite image resolutions and noise. One possibility is to assume that the images are small compared to the scale over which the lens properties change, so that the entries of the transformation matrix are constant (\cite{bib:Tessore2}).

As two examples to test the range of validity of this approximation and show differences between the choice of variables, we simulate typical NFW profiles with masses $M = 10^{15} M_\odot$ and $M = 0.5 \cdot 10^{15} M_\odot$, and a concentration $c_\mathrm{NFW} = 3$ (s.\ \cite{bib:Merten} for example cluster reconstructions in this parameter range) and SIEs with axis ratios $f = 0.9$ and $f = 0.8$ in the notation of \cite{bib:Kormann}. Sources are placed at increasing distance from the fold and cusp points in the source plane and for the cusp, the sources are placed on the symmetry axis connecting the lens centre with the cusp. 

For both profiles, we investigate the goodness of approximation of Equation~\eqref{eq:limit} dependent on the distance to the critical curve when using convergence and reduced shear and when using potential derivatives. As the distance to the critical curve is no observable, we plot the ratios of convergences and potential derivatives with respect to the relative image distance. Figure~\ref{fig:accuracy_comparison} shows the results using $\kappa$ as in \cite{bib:Tessore}, Figure~\ref{fig:accuracy_comparison_phi} the results using $\phi_{11}$ as in \cite{bib:Wagner, bib:Wagner2}. For both profiles, image $A$ is the one with positive parity, image $B$ the one with negative parity. In the cusp case, image $B$ is the one closest to the cusp.

From a comparison of Figure~\ref{fig:accuracy_comparison} with \ref{fig:accuracy_comparison_phi}, we deduce that both choices of parametrisation are equally well-suited for fold configurations of SIEs, the ansatz using the convergence excels over the one with potential derivatives for cusp configurations of SIEs and vice versa for fold configurations of NFW profiles. The very small decreasing slope for the NFW profile (right plot in Figure~\ref{fig:accuracy_comparison_phi}) explains the high accuracy of the Taylor approximation that we found in \cite{bib:Wagner2}. (For that case, the quadrupole moments have to be used as observables because the system of equations using the transformation matrix is underdetermined.)

We note that ratios of $\phi_{11}$ smaller than one for an SIE cusp configuration are achieved when interchanging $A$ and $B$ with respect to the choice of Figure~\ref{fig:accuracy_comparison}, as $\phi_{11}^{(B)} > \phi_{11}^{(A)}$, while at the folds of both profiles $|\phi_{11}^{(B)}| < |\phi_{11}^{(A)}|$ holds.

The precision of the ratios of potential derivatives is given by the precision of the transformation and the relative distances between the centres of light of the images. As an example, \cite{bib:Gorenstein} obtain the eigenvalues of $T$ with uncertainties below 5\% and the direction of the semi-major axis of $T$ with an uncertainty of 7\% in a $\chi^2$-parameter estimation mapping image $A$ in Q0957+561 onto image $B$ in VLBI observations. Compared to these uncertainties for the $T_{ij}$, the relative uncertainty of the distance between $A$ and $B$ on the order of $\mathcal{O}(10^{-5})$ is negligible.
%and compare the precision to which the reduced shear components can be determined with the precision achievable with the quadrupole moments. 

%%%%%%%%%%%%%%%%%%
\section{Implementation}
\label{sec:implementation}

In this work, we identify resolved features in the intensity distributions of the images manually and leave automated and more advanced techniques to further work.

The transformation matrix has four entries and thus four degrees of freedom that amount to a rotation angle, two stretching factors and a parity inversion as physical degrees of freedom. Hence, we need at least three positions, i.e.\ two linearly independent vectors, in each image that can be mapped to their counterparts in another image to determine the transformation matrix. 

In the vicinity of a fold in the leading order approximation we consider, two points, i.e.\ one vector in each image, suffice because there is always a parity inversion that is fixed and the only degrees of freedom are the rotation  angle and one stretching factor to change the area and thus determine the absolute value of the magnification ratio caused by gravitational lensing. 

If more than the necessary amount of points is given, the system of equations is overconstrained. If the necessary amount of images is given, the most efficient way to solve for the lens parameters is to parametrise the entries of the transformation matrices by the right-hand sides of Equations~\eqref{eq:T11} to \eqref{eq:T22} or Equations~\eqref{eq:T11_phi} to \eqref{eq:T22_phi} and insert them into the system of transformation equations
\begin{equation}
\left(\boldsymbol{x}_{I\alpha} -\boldsymbol{x}_I\right) = T^{(A,I)} \left( \boldsymbol{x}_{A\alpha} - \boldsymbol{x}_A \right) \;, \quad I = B, C, ... \;, \quad \alpha = 1,2,...
\label{eq:system_of_equations}
\end{equation}
in which $\boldsymbol{x}_{I\alpha} \in \mathbb{R}^2$ denote the reference points of the individual images to which the reference points in the reference image $\boldsymbol{x}_{A\alpha} \in \mathbb{R}^2$ are mapped. $\boldsymbol{x}_I$ are their the centres of light and $\boldsymbol{x}_A$ the centre of light of the reference image. This system of equations is solved for the lens parameters in $T^{(A,I)}$ as a non-linear least squares parameter estimation problem (NLLSP), minimising the deviations between the right-hand side and the left-hand side. To account for different, possibly correlated uncertainties of the reference point positions in each image, the deviations are weighted by the inverse variance of these uncertainties. 

To determine the higher order ratios of potential derivatives at the critical points, we further replace the second order potential derivatives by Equations~\ref{eq:rel_phis1} -- \ref{eq:rel_phis4}.

Subsequently, the resulting lens parameters and their covariances are taken as input for an importance sampling Monte-Carlo simulation. The median lens parameter values of all samples and their $1-\sigma$ confidence intervals (simply denoted as confidence intervals) are the final result. The number of samples is chosen to be at least 10~000 to assure the weights of the importance sampling to be well-balanced. 

If more than the necessary amount of images is given, we do not parametrise $T$ by means of the lens parameters, but employ a different, more efficient parametrisation for $T$ in the system of Equations~\ref{eq:system_of_equations} that reduces the non-linearity of the NLLSP. This implementational detail will be further described in \cite{bib:Tessore2} and is used to determine the lens parameters in Section~\ref{sec:f_g_complete}.

%%%%%%%%%%%%%%%%%%
\section{Example application}
\label{sec:examples}

\begin{figure*}[t!]
\centering
 \raisebox{2.35cm}{\includegraphics[width=0.4\textwidth]{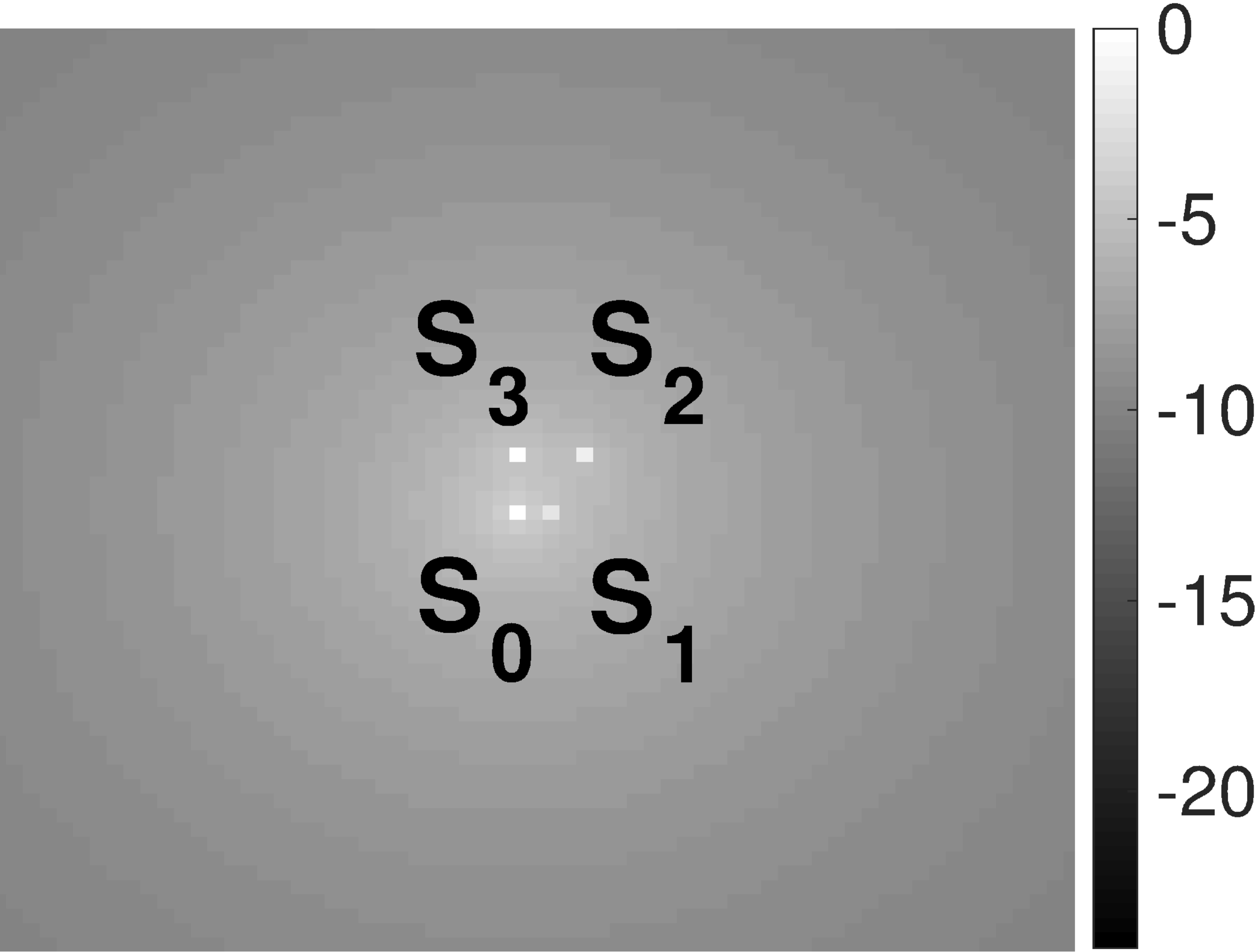}}
    \includegraphics[width=0.48\textwidth]{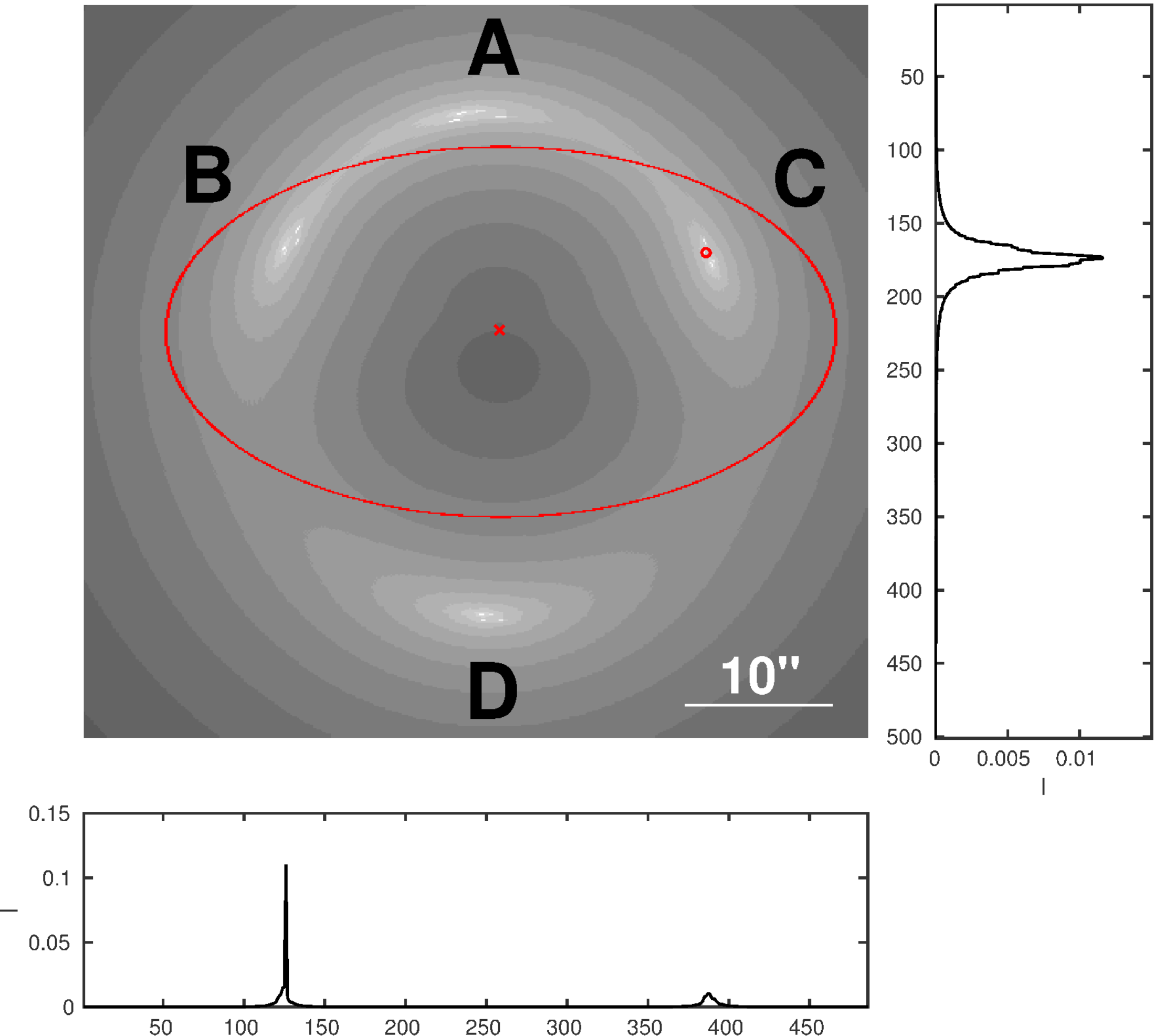}
    \caption{Logarithmic intensity distribution of a source consisting of four Sérsic profiles, as detailed in Section~\ref{sec:sources}, simulating a galaxy with resolved brightness features (left). Multiple images of this source generated by the SIE lens introduced in \cite{bib:Wagner2} and described in Section~\ref{sec:lens} (right). The lens centre is highlighted by a red cross, the red circle marks the point through which the horizontal and vertical intensity profiles run that are shown beneath and right of the picture.} 
    \label{fig:sie_simulation}
\end{figure*}

%%%%%%%%%%%%%%%%%%
\subsection{Simulation of sources}
\label{sec:sources}

In order to set up the transformation matrices between the images by the reference matching described in Section~\ref{sec:implementation} and estimate their uncertainties, we simulate a source galaxy as a superposition of four round Sérsic profiles with Sérsic index $n=4$ to obtain four reference points in each image. 
The first source consists of a Sérsic profile with $r_\mathrm{e} = 15~\mathrm{px}$, denoted as $S_0$, and three adjacent Sérsic profiles with $r_\mathrm{e} = 5~\mathrm{px}$, denoted as $S_1, S_2, S_3$, at offsets $(2~\mathrm{px}, 0~\mathrm{px})$, $(4~\mathrm{px}, 4~\mathrm{px})$, and  $(0~\mathrm{px}, 4~\mathrm{px})$ from the centre of the first one. For an easier identification of the points, we scale the intensity distributions of the three smaller Sérsics to 0.1, 0.3, and 0.7 of the large one. Figure~\ref{fig:sie_simulation} (left) shows the logarithmic intensity profile for this source.

The second source has the same Sérsic profiles as the first one, but the three smaller Sérsic profiles are at offsets $(3~\mathrm{px}, 0~\mathrm{px})$, $(5~\mathrm{px}, 5~\mathrm{px})$, and  $(0~\mathrm{px}, 5~\mathrm{px})$. Hence, the reference points in the second source are farther away from each other, so that the accuracy of the local lens properties dependent on the distance between the reference points within the images can be investigated. 

%%%%%%%%%%%%%%%%%%
\subsection{Simulation of the lens}
\label{sec:lens}

As lens, we use the SIE introduced in \cite{bib:Wagner2}, as it is the simplest lens for which all discussed image configurations can be analysed. Figure~\ref{fig:sie_simulation} (right) shows the images generated by the SIE for the first source described in Section~\ref{sec:sources}. Placing the second source at the same position in the source plane,  the semi-major and semi-minor axes of the quadrupole moments of its multiple images are about 30\% longer than the axes of the images of the first source. The image configuration $(A, B, C)$ is an asymmetric cusp configuration in which the central image $A$ has positive parity. (In \cite{bib:Wagner2} we already analysed symmetric cusp configurations with a central image of negative parity in detail.)

% reality check
The extensions of these simulated multiple images match the scale of observed ones and, as already stated in \cite{bib:Wagner2}, the semi-major axis of the critical curve of the SIE has an extension of 23'' and is thus similar to observed effective Einstein radii and the effective Einstein radius of the simulated HERA cluster, \cite{bib:Meneghetti}.

%%%%%%%%%%%%%%%%%%
\subsection{Second order lens parameters of all images}
\label{sec:f_g_complete}

We first determine the ratios of convergences and reduced shear components from the observables of all multiple images from the first source by employing the algorithm outlined in Section~\ref{sec:implementation}. All weights in the NLLSP optimisation are set to 1, 10~000 samples are used for the Monte-Carlo simulation and image $A$, the positive parity image closest to the cusp, is chosen as reference image. 

\begin{figure*}[p]
\centering
  \includegraphics[width=0.49\textwidth]{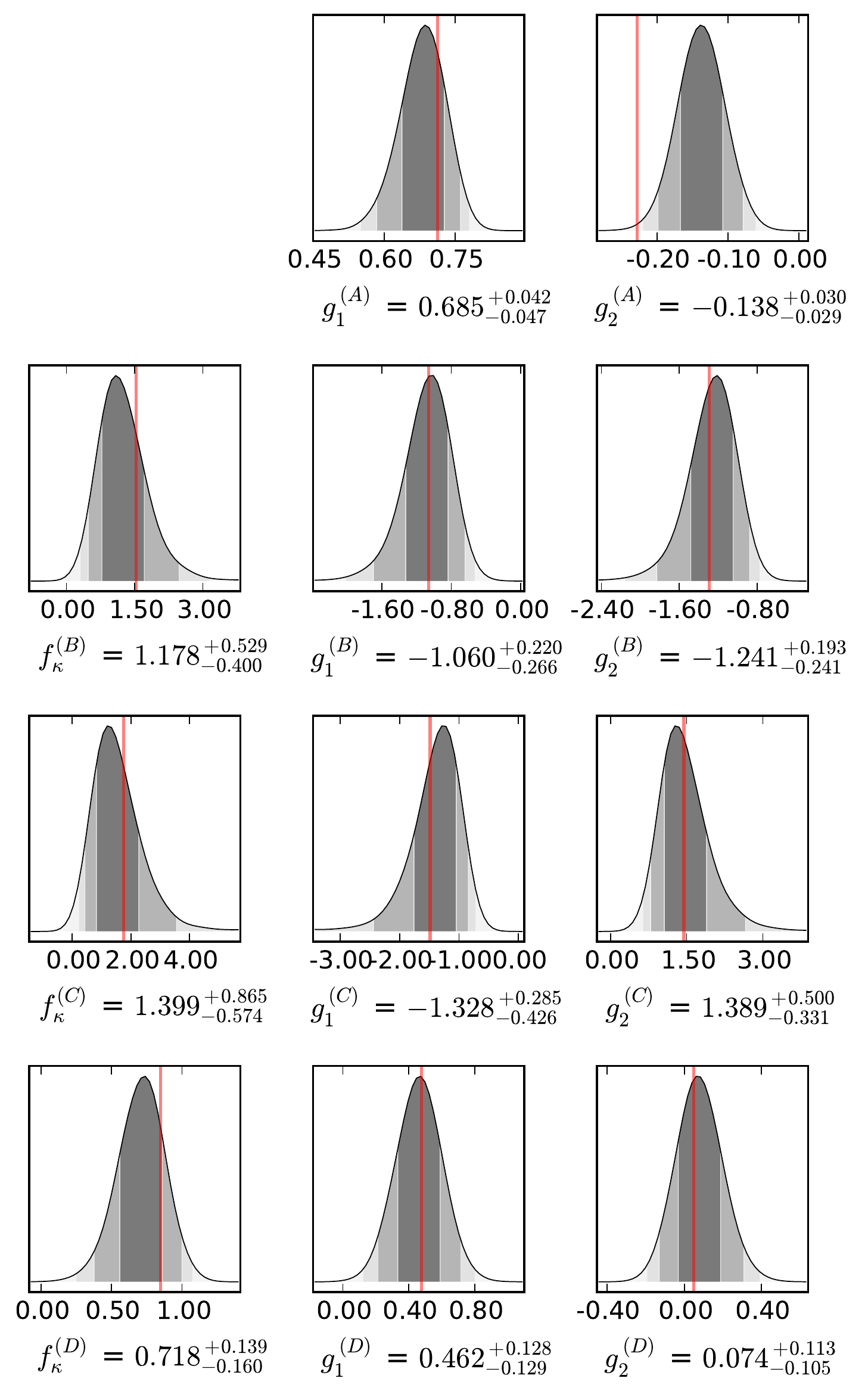}
    \caption{Comparison of the probability density distributions of the convergence ratios $f_\kappa$ and reduced shear components $g_{i}$, $i=1,2$, of the SIE lens at the centres of light of the four multiple images shown in Figure~\ref{fig:sie_simulation} with their true values (red lines). The dark grey-shaded area, the grey-shaded area, and the light grey-shaded area delimit the $1-$, $2-$, and $3-\sigma$ confidence intervals, respectively.} 
    \label{fig:f_g_complete}
\end{figure*}

Figure~\ref{fig:f_g_complete} shows the resulting probability density distributions of the $f_\kappa$- and $g_i$-values, $i=1,2$, for all four multiple images at their centres of light. The confidence intervals for $g_{1}^{(A)}$ and $g_{2}^{(A)}$ are smaller than the ones for the remaining lens parameters because all remaining $f_\kappa$ and $g_i$ depend on $g_{1}^{(A)}$ and $g_{2}^{(A)}$.

To compare with the true $f_\kappa$ and $g_i$ at the centre of light, we plot the true value on top of each probability density distribution. We observe that all true values are within the confidence intervals, except for $g_{2}^{(A)}$.

%%%%%%%%%%%%%%%%%%
\subsection{Second order lens parameters at a fold}
\label{sec:fold}

\begin{figure*}[p]
\centering
  \includegraphics[width=0.3\textwidth]{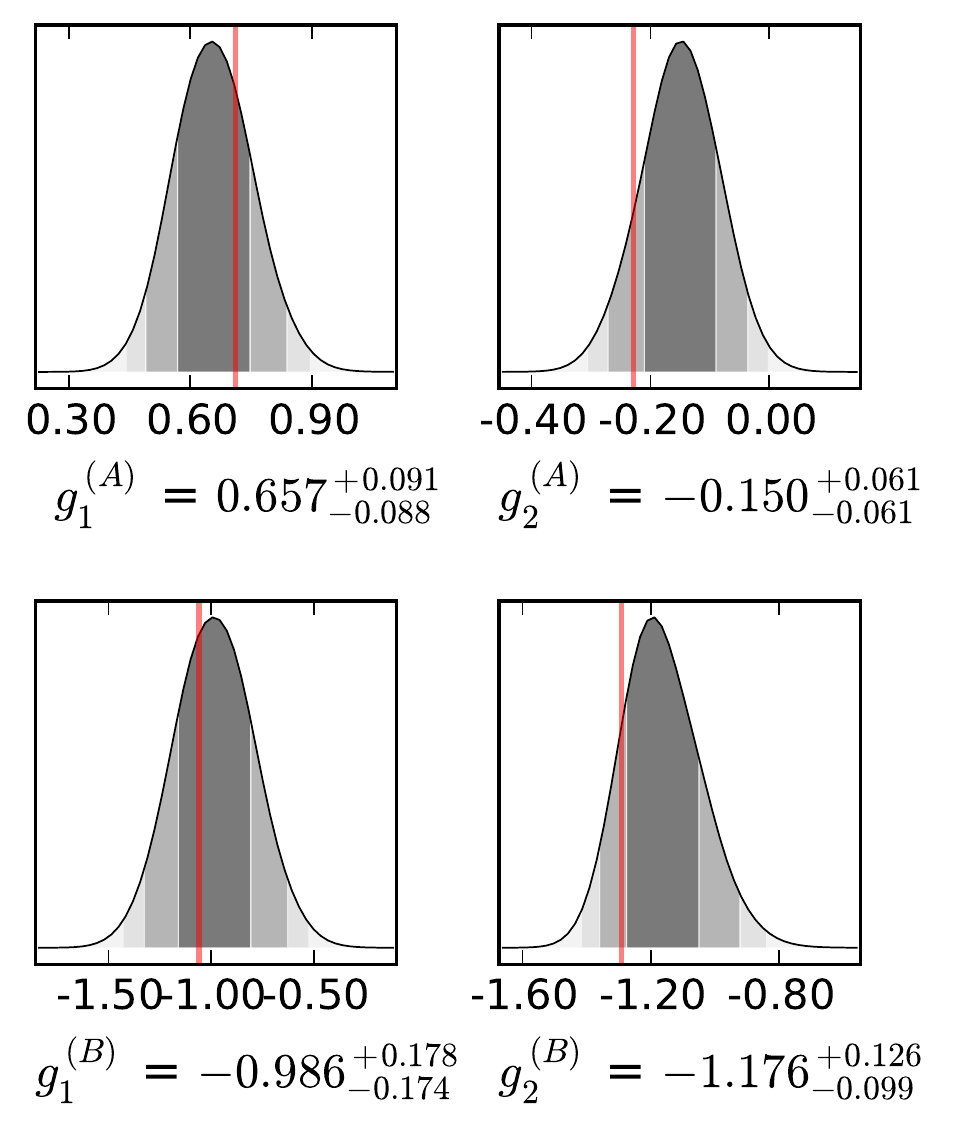}
    \includegraphics[width=0.3\textwidth]{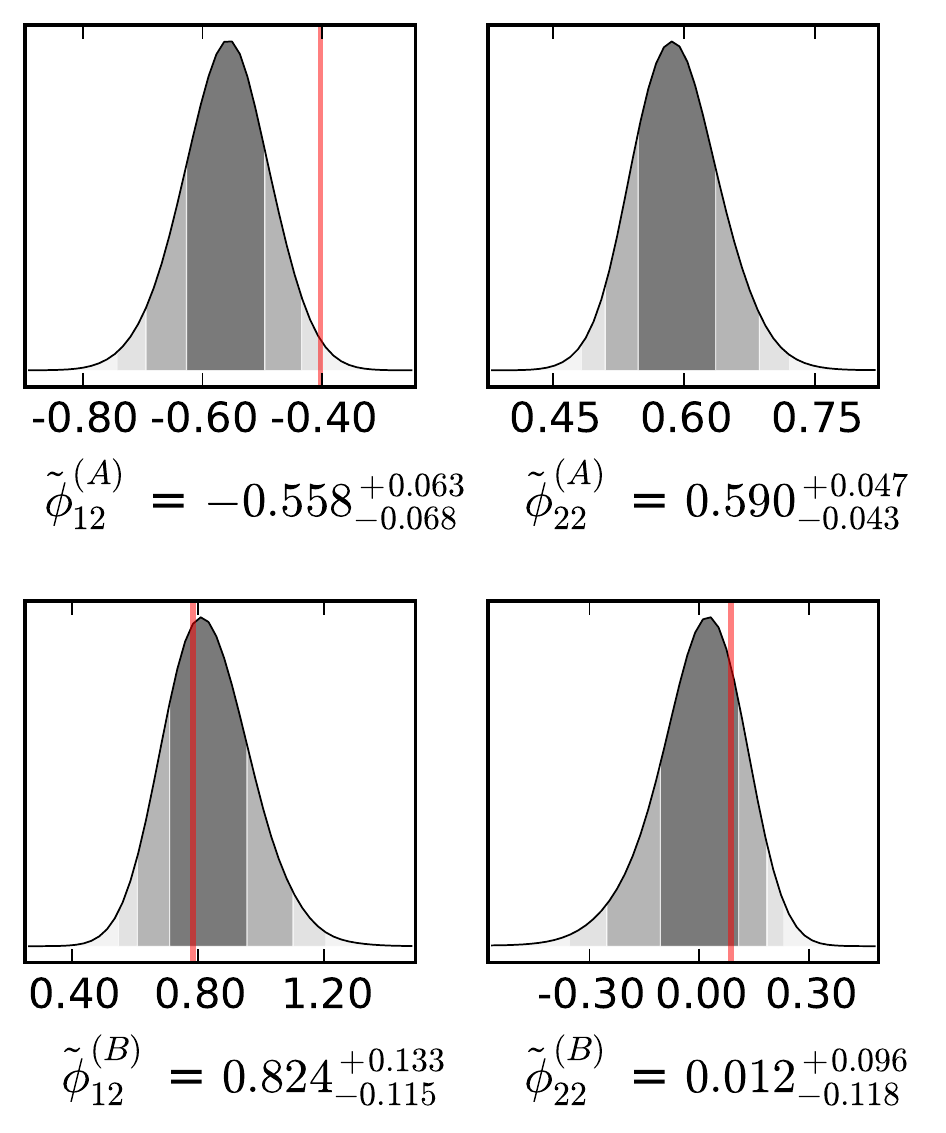}
    \caption{Comparison of the true lens parameters (red lines) at a fold with the probability density distributions of the lens parameters determined by our approach in the parametrisation of \cite{bib:Tessore} (first two columns) and in the parametrisation of \cite{bib:Wagner2} (last two columns).} 
    \label{fig:fold}
\end{figure*}

Considering the image pair $(A,B)$ in Figure~\ref{fig:sie_simulation} as a fold configuration, we determine the reduced shear components under the assumption that $f_\kappa \approx 1$ and for the same implementational specifications as described in Section~\ref{sec:f_g_complete}. As can be read off the first two colums in Figure~\ref{fig:fold}, the true $g_1$ at the centre of light lie within the confidence intervals of both images, while the $g_2$ lie outside. 

Subsequently, we calculate probability density distributions of the ratios of second order potential derivatives and plot them in the last two columns in Figure~\ref{fig:fold}. This parametrisation leads to worse results compared to the one using the reduced shear components because the true ratios of second order potential derivatives for image $A$ are farther away from the most likely ratios of second order potential derivatives obtained by our approach. $\tilde{\phi}_{22}^{(A)} = 0.367$ is not even contained in the $3-\sigma$ confidence interval that limits the plot range.

Calculating $f_\kappa^{(B)} = 1.53$ and $f_\phi^{(B)} = 1.21$, we find that both values have a large deviation from Equation~\ref{eq:limit}, which explains why the reconstructed lens parameters show high inaccuracies for both parametrisations. In addition, we conclude that for increasing ellipticity of the SIE the parametrisation using the convergence ratios and reduced shears yields more accurate results than the one using the ratios of the potential derivatives.

%%%%%%%%%%%%%%%%%%
\subsection{Third order lens parameters at a fold point}
\label{sec:fold_cc}

\begin{figure}[h]
\centering
    \includegraphics[width=0.3\textwidth]{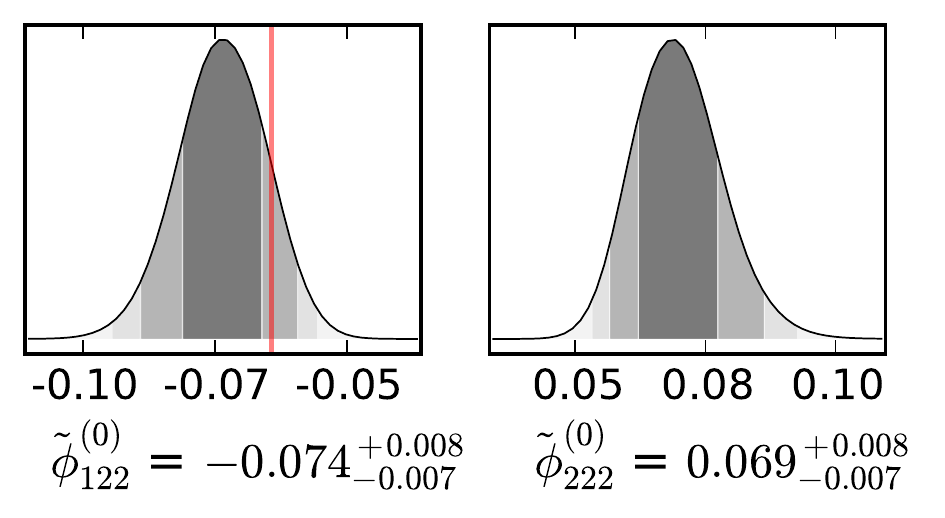}
    \caption{Comparison of the true third order ratios of potential derivatives (red lines) at a fold with the probability density distributions of the ratios of potential derivatives determined by our approach.} 
    \label{fig:fold_cc}
\end{figure}

Keeping the implementational specifications as described in the previous sections, we determine the probability density distributions for the third order ratios of potential derivatives as shown in Figure~\ref{fig:fold_cc}. As expected from the results of Section~\ref{sec:fold}, the true values of the ratios of potential derivatives at the fold point lie outside the confidence intervals -- the true value of $\tilde{\phi}_{222}^{(0)}$ lies beyond the plot range -- and the reconstruction of the critical curve in the vicinity of the fold will consequently be very inaccurate.

%%%%%%%%%%%%%%%%%%
\subsection{Higher order lens parameters at a cusp point}
\label{sec:cusp_cc}

The accuracy of the reconstruction of the critical curve in the vicinity of the cusp is summarised in Figure~\ref{fig:cusp_cc}, employing the same implementational specifications as before. The first two columns are obtained using the image pair $(A,B)$, the last two columns are obtained using the image pair $(A,C)$ to determine the ratios of potential derivatives and the centre of light position of image $A$. From the first row, we conclude that the accuracy of the slope of the parabola that approximates the critical curve in the vicinity of a cusp is very low because the true values of the third and fourth order ratios of potential derivatives lie far off the $3-\sigma$ confidence intervals. Contrary to that, the true value of $x_{A1}$, centre of light position of image $A$, lies within the confidence interval. Using image pair $(A,B)$, we find that $x_{A2}$ lies within the confidence interval as well, so that the true cusp position is determined in more than 68\% of the cases.

%%%%%%%%%%%%%%%%%%
\subsection{Comparison to the quadrupole moment as observable}
\label{sec:quadrupole}

To compare the results obtained by means of the transformation matrix with the results retrievable from the quadrupole moment as observable, we have to specify uncertainties for the quadrupole moment observables, the axis ratio $r_I$ and the orientation angle of the semi-major axis $\varphi_I$, $I=A,B,C$, and generate the probability density distributions for the lens parameters. As first estimate, we assume that all uncertainties in the observables follow a Gaussian distribution and are uncorrelated. The distance between the images is usually determined to high precision (see Section~\ref{sec:errors}), so that a standard deviation of $10^{-4}\,\delta_{IJi}$, $i=1,2$, for the coordinatewise distances between the images seems appropriate. For the axis ratio, we estimate a standard deviation of $0.1 \, r_I$ and for the orientation angle $\pm 5^\circ$ from visual inspection of some HST observations. From these normal distributions of observables that are obtained for the first source in our SIE simulation, we generate 10~000 samples and determine the probability density distributions for the ratios of potential derivatives.

\begin{figure*}[p]
\centering
    \includegraphics[width=0.28\textwidth]{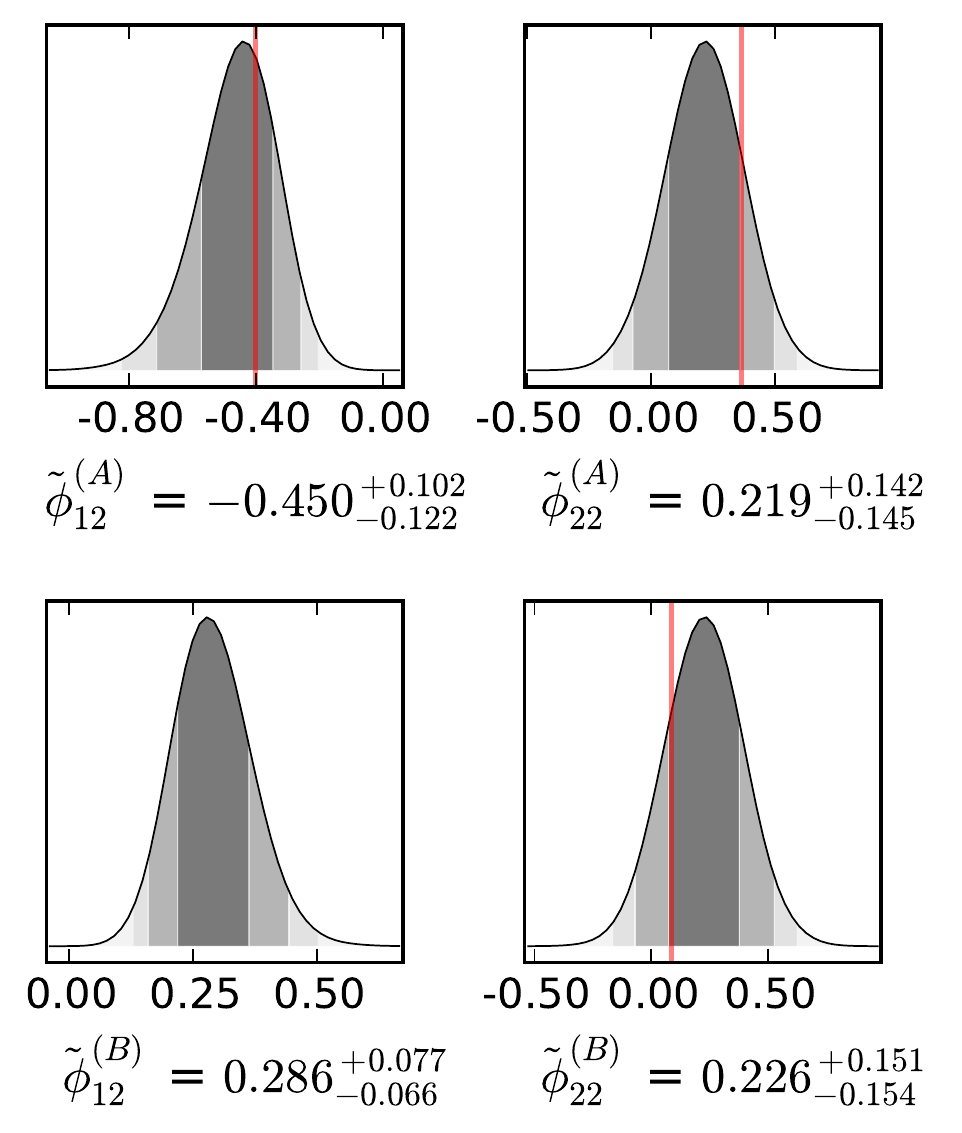}
        \includegraphics[width=0.28\textwidth]{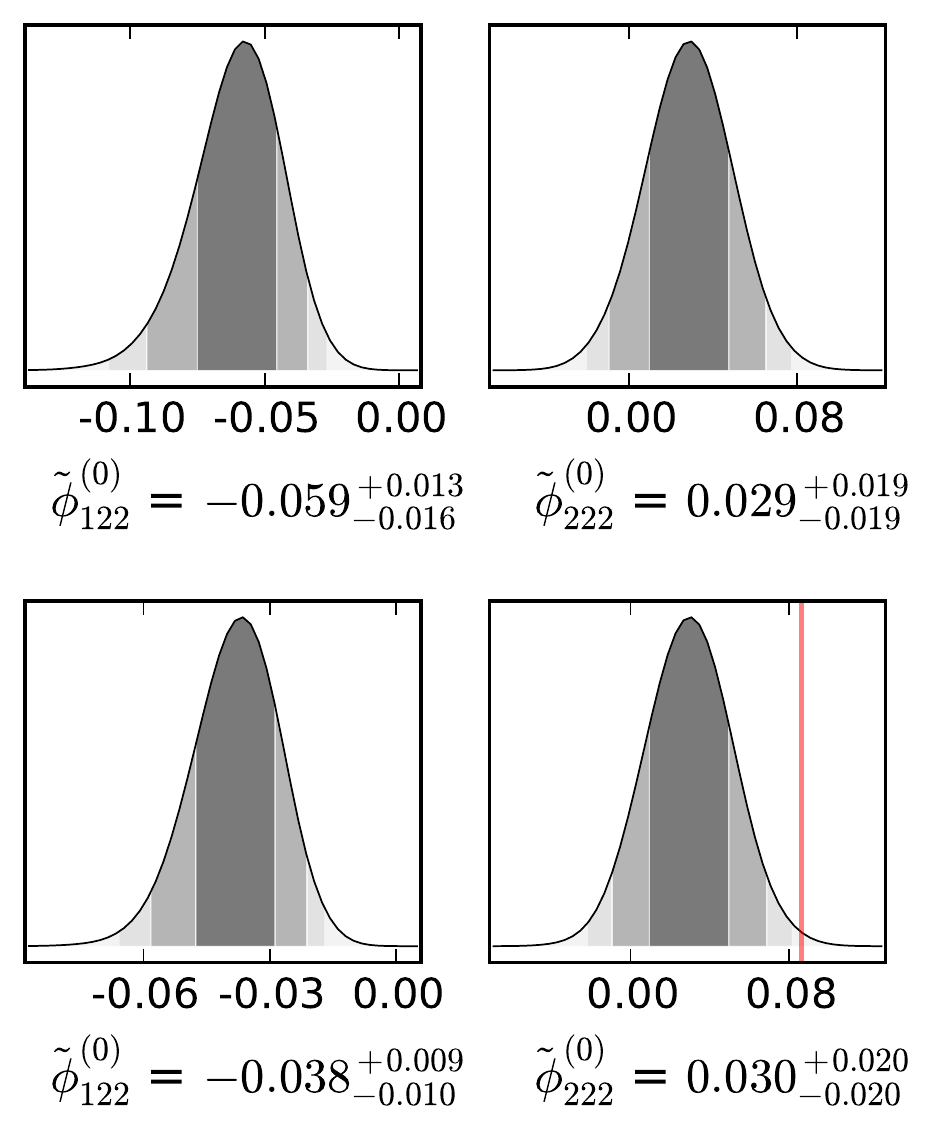}
    \caption{Comparison of the true ratios of potential derivatives (red lines) at a fold with the probability density distributions of the ratios of potential derivatives determined by our approach using the quadrupole moment as observable instead of the transformation matrix. The first row of the last two columns uses image $A$ and the bottom row uses image $B$ for the reconstruction of the third order potential derivatives, respectively.} 
    \label{fig:fold_quadrupole}
\end{figure*}

The results for the second order ratios of potential derivatives at a fold are plotted in the first two colums of Figure~\ref{fig:fold_quadrupole}\footnote{To obtain these results for $\tilde{\phi}^{(0)}_{222}$, the approximation that the images are extended orthogonally to the critical curve is used, as detailed in \cite{bib:Wagner2}}. They show that employing the quadrupole yields slightly higher probabilities to retrieve the true values of the second order potential derivatives, compared to using the transformation matrix (see the last two columns in Figure~\ref{fig:fold}). However, the ratios of third order potential derivatives, shown in the last two columns of Figure~\ref{fig:fold_quadrupole}, have larger confidence intervals and a very low probability to retrieve the true values. Therefore, using the quadrupole as observable yields worse results than employing the transformation matrix for the reconstruction of the critical curve at the fold.

\begin{figure*}[p]
\centering
    \includegraphics[width=0.28\textwidth]{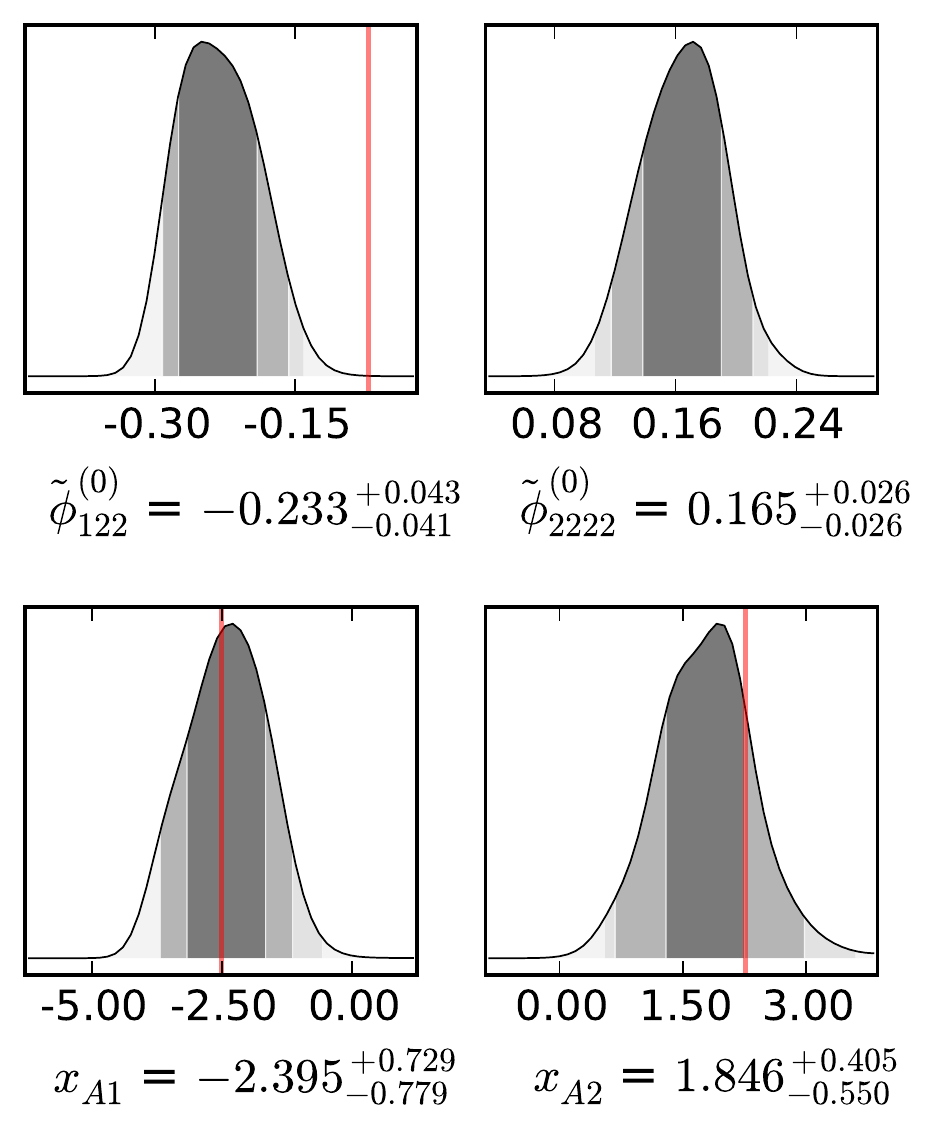}
        \includegraphics[width=0.28\textwidth]{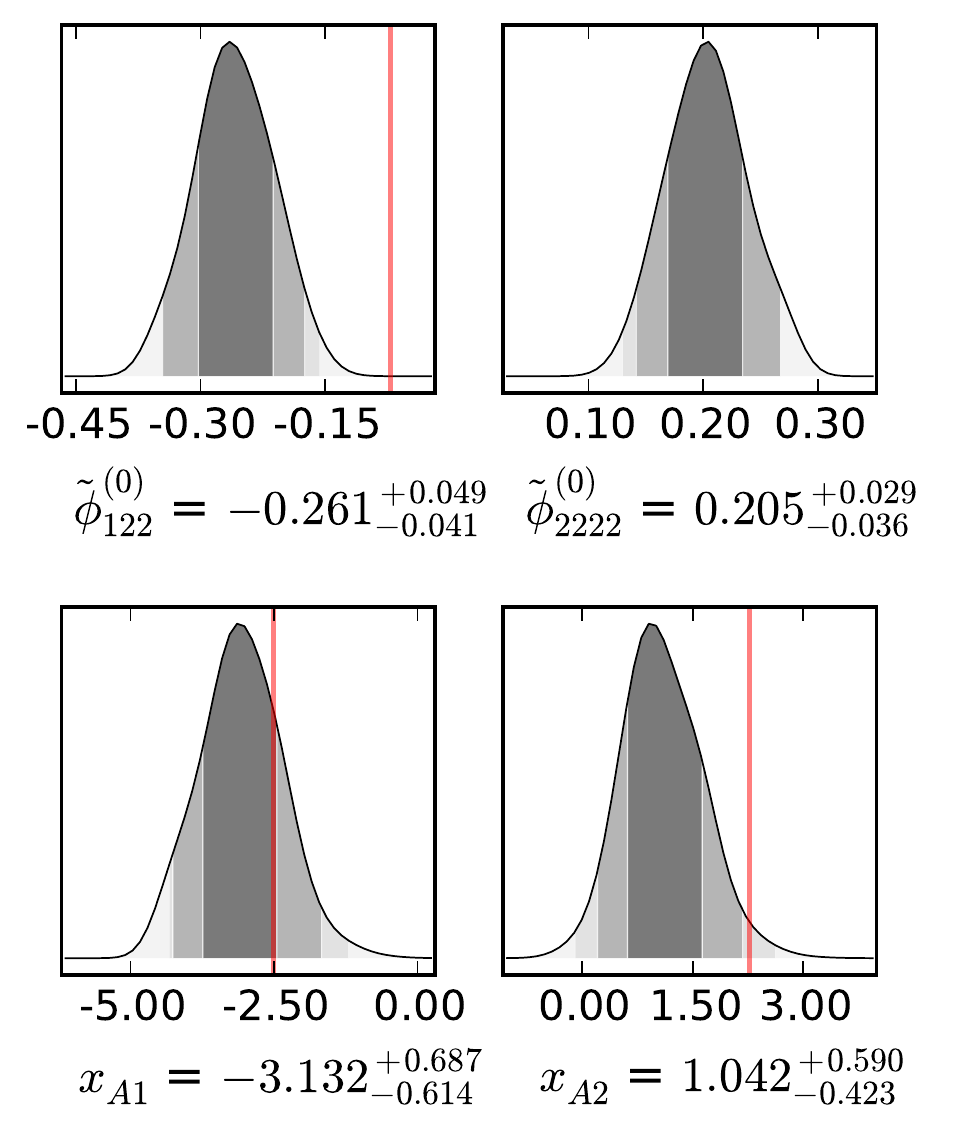}
    \caption{Comparison of the true ratios of potential derivatives and image positions (red lines) at a cusp with the probability density distributions of the ratios of potential derivatives and image positions determined by our approach. The first two columns use the image pair $(A,B)$ and the last two columns the image pair $(A,C)$ for the reconstruction, respectively.} 
    \label{fig:cusp_cc}
\end{figure*}  

\begin{figure*}[p]
\centering
    \includegraphics[width=0.28\textwidth]{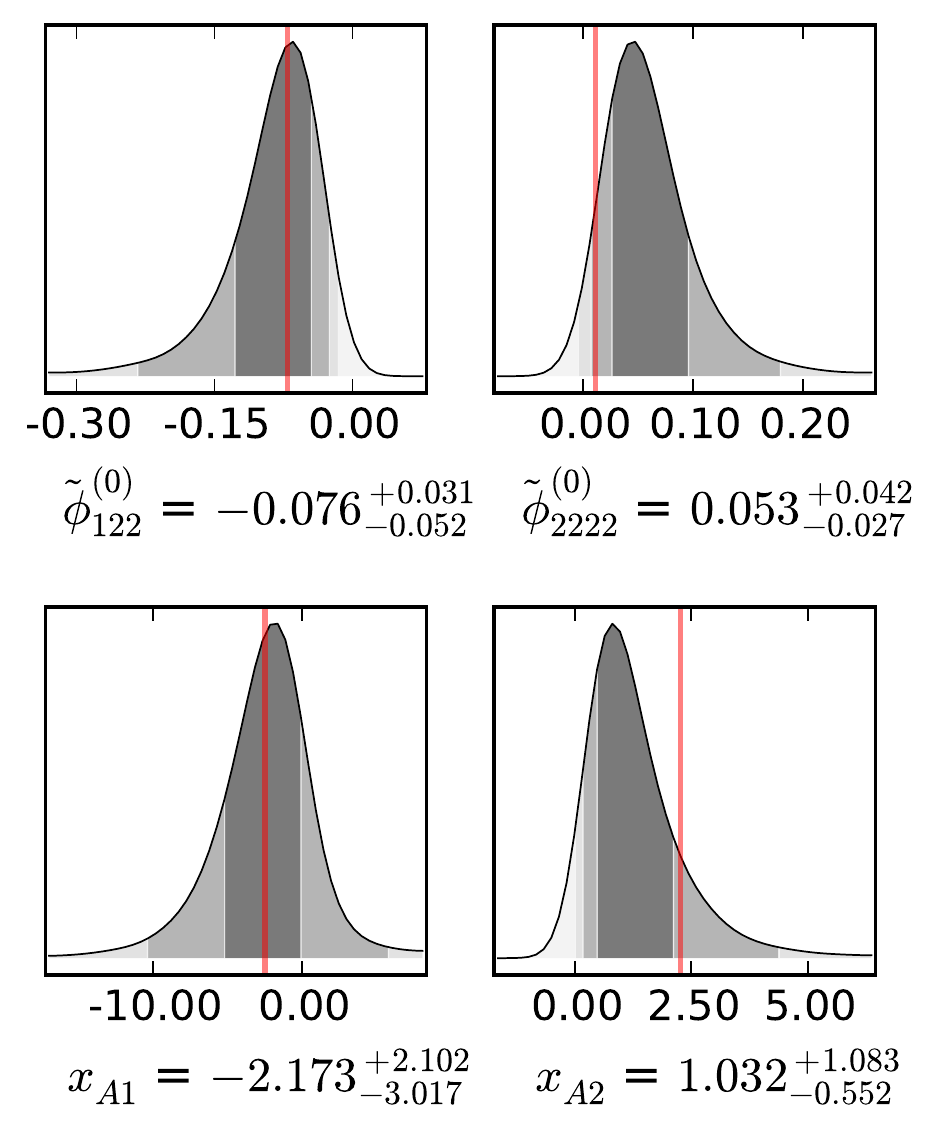}
        \includegraphics[width=0.28\textwidth]{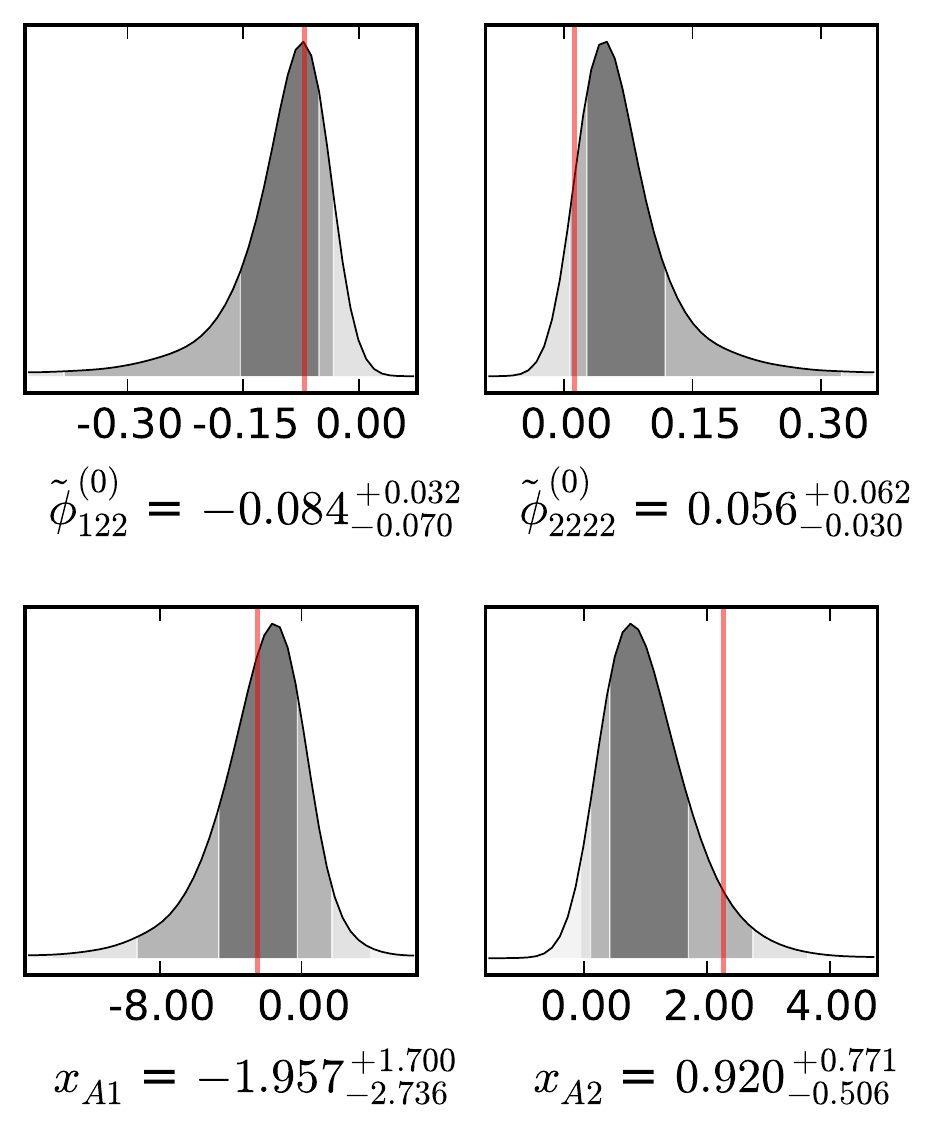}
    \caption{Comparison of the true ratios of potential derivatives and image position $A$ (red lines) at a cusp with the probability density distributions of the ratios of potential derivatives and image position $A$ determined by our approach using the quadrupole moment as observable instead of the transformation matrix. The first two columns use the image pair $(A,B)$ and the last two colums use the image pair $(A,C)$ for the reconstruction, respectively.} 
    \label{fig:cusp_quadrupole}
\end{figure*}

\begin{figure*}[hp!]
\centering
  \includegraphics[width=0.75\textwidth]{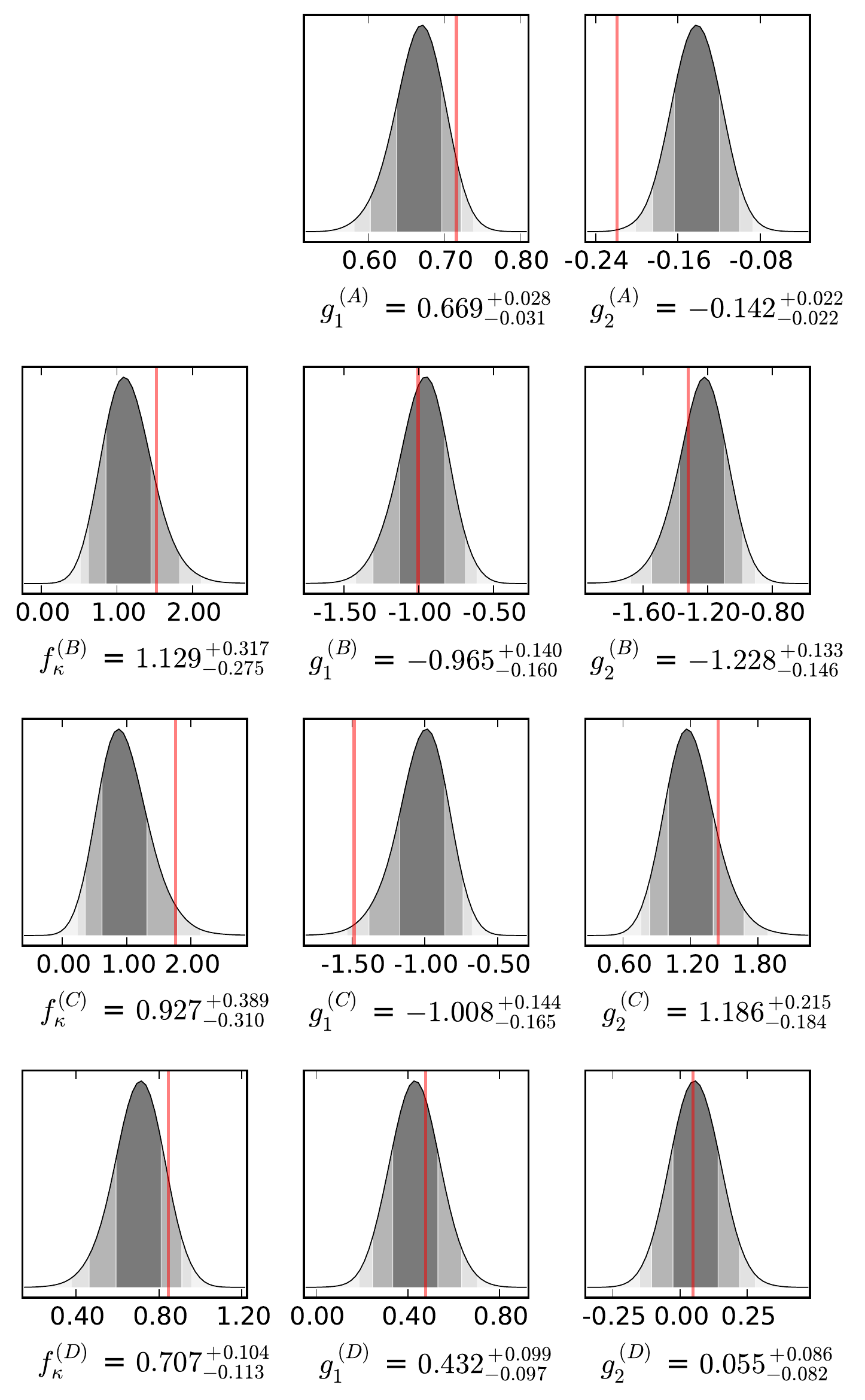}
    \caption{Comparison of the probability density distributions of the convergence ratios $f_\kappa$ and reduced shear components $g_{i}$, $i=1,2$, of the SIE lens at the centres of light of the four multiple images of the second source with their true values (red lines). The dark grey-shaded area, the grey-shaded area, and the light grey-shaded area delimit the $1-$, $2-$, and $3-\sigma$ confidence intervals, respectively.} 
    \label{fig:f_g_complete_larger_image}
\end{figure*}

Analogously, we determine the ratios of potential derivatives at the cusp point and the position of image $A$ once from the image pair $(A,B)$, once from the image pair $A,C)$ using the quadrupole moments of the images as observables. The results are plotted in the first and the last two columns of Figure~\ref{fig:cusp_quadrupole}, respectively. As for the transformation matrix, the image pair $(A,B)$ has the highest probability of retrieving the true values for the ratios of potential derivatives and $\boldsymbol{x}_A$. Comparing Figure~\ref{fig:cusp_quadrupole} with Figure~\ref{fig:cusp_cc}, using the quadrupole moment clearly yields higher probabilities for the true values than employing the transformation matrix. Yet, this comes at the cost of broader confidence intervals. Another disadvantage of the quadrupole moment as observable is that it may still contain a bias due to the intrinsic ellipticity of the source.

%%%%%%%%%%%%%%%%%%
\subsection{Influence of the image size}
\label{sec:image_size}

To investigate the influence of the size of the multiple images on the accuracy of the reconstruction, we repeat the analysis of Section~\ref{sec:f_g_complete} for the second source. Comparing the plots in Figure~\ref{fig:f_g_complete} with the ones in Figure~\ref{fig:f_g_complete_larger_image}, we observe that the confidence intervals in the former are larger, so that the true parameter values have a higher probability to be correctly determined for the smaller source. The reason for the decreasing accuracy when increasing the image extensions is the decreasing validity of the assumption that the entries of the magnification matrix are constant over the extension of each multiple image. The semi-major axis of the quadrupole moment of image $A$ is 9\% of the distance between image $A$ and $B$ for the first source and 12\% for the second, which gives an estimate at which scales higher order moments should be taken into account.

%%%%%%%%%%%%%%%%%%
\subsection{Influence of noise}
\label{sec:noise}

\begin{figure*}[hp!]
\centering
  \includegraphics[width=0.75\textwidth]{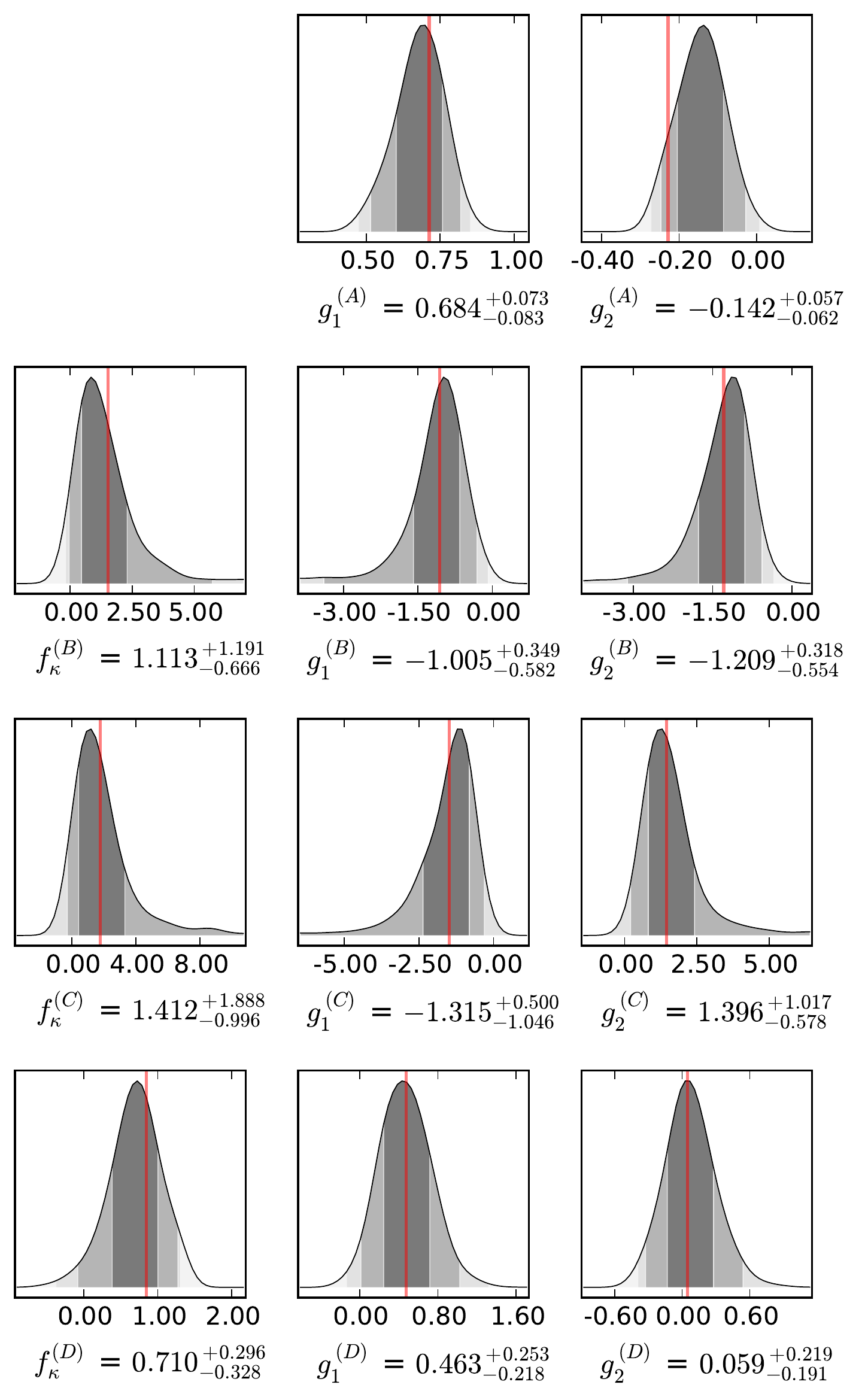}
    \caption{Comparison of the probability density distributions of the convergence ratios $f_\kappa$ and reduced shear components $g_{i}$, $i=1,2$, of the SIE lens at the centres of light of the four multiple images of the first source with their true values (red lines) with an increased uncertainty in the reference positions of 2 px. The dark grey-shaded area, the grey-shaded area, and the light grey-shaded area delimit the $1-$, $2-$, and $3-\sigma$ confidence intervals, respectively.} 
    \label{fig:noise1}
\end{figure*}

Taking into account the noise in the observations, the reference points may not be accurately determined at the pixel-level, even when setting them manually. Keeping all other (implementational) specifications as before, we equally increase the uncertainty in all reference point positions from 1 px to 2 px, while assuming that they are still uncorrelated. In addition, we have to increase the number of samples to 200~000 because the number of effective samples in the importance sampling step falls below 100 samples otherwise. This leads to significant multi-modal artefacts in the probability density distributions.

Comparing the resulting lens parameter probability density distributions shown in Figure~\ref{fig:noise1} with the ones in Figure~\ref{fig:f_g_complete}, we observe that the confidence intervals are increased such that the true lens parameter values are more often correctly retrieved. Giving the reference point positions in different images different uncertainties, we observe that the confidence intervals of the respective lens parameters are broadened accordingly. Yet, the confidence intervals can be as large as the parameter value itself, especially for the $f_\kappa$, so that the approach becomes unreliable for increasing noise.

%%%%%%%%%%%%%%%%%%
\section{Conclusion}
\label{sec:conclusion}

We investigated which model-independent information about a gravitational lens can be gained by linearly mapping multiple images with clearly resolved features onto each other close to fold or cusp critical points and how accurately these properties can be determined. As results, the following can be stated:
\begin{itemize}
\item The approach developed in \cite{bib:Tessore} can be reparametrised in terms of ratios of potential derivatives to show that it yields the same information about the magnification matrices in the vicinity of the critical curve as the approach developed in \cite{bib:Wagner2}. 
\item The approach of \cite{bib:Tessore} determines the magnification matrices up to a scale factor for all multiple images (i.e.\ convergence ratios and reduced shear components) at the image positions but it cannot be applied to images generated by axisymmetric lenses. 
\item The approach of \cite{bib:Wagner2} determines the magnification matrices up to a scale factor at the image positions and an approximation to the critical curve in the vicinity of the images close to a fold or a cusp critical point. But it cannot determine the lens properties at the position of the counter image lying on the opposite side of the lens centre or the position of the central image close to the lens centre. 
\item Both approaches in their current implementation assume that the convergence and shear variations are negligible over the extensions of the individual multiple images, so that images can be mapped onto each other using a linear transformation. 
\item Combining both approaches yields model-independent information about the magnification matrices at the positions of all multiple images, their relative parities, and allows to reconstruct the critical curve in the vicinity of fold and cusp critical points.
\item We simulate a galaxy-cluster scale SIE lens and a source galaxy consisting of four elliptical Sérsic profiles to mimick a galaxy that shows clearly resolved features in its intensity distribution. In each image, we need to identify at least three distinctive features from which we then assume that they are mapped onto each other by the transformation matrix.  As we intend to analyse the five-image configuration of a source at redshift $z_\mathrm{s}=1.675$ in the galaxy cluster CL0024+1654 next, the scales and extensions of our simulated lens and source are chosen to be similar to that observation. Hence, the following results from analysing the simulated multiple-image configurations can be used as a calibration for the observational case. 
\item A comparison of the parametrisations from \cite{bib:Tessore} and \cite{bib:Wagner2} shows that the symmetric approach using ratios of convergences and reduced shear components is more robust than the one using ratios of potential derivatives because the former does not require a transformation into a special coordinate system. Furthermore, it yields higher probabilities to retrieve the true magnification matrices up to a scale factor.
\item A comparison between using the transformation matrix between pairs of images to retrieve the lens parameters and using the quadrupole moments (characterised by their axis ratios and orientation angles, as detailed in \cite{bib:Wagner2}) of the individual images reveals that the approach using the quadrupole moments yields higher probabilities to retrieve the true lens parameters for the cusp and yields lower probabilities to accurately retrieve the lens parameters for the fold. 
\item The approximation that convergence and shear are constant over the image areas becomes inaccurate when the semi-major axis of the quadrupole moment of the images is on the order of 10\% of the relative distance between the centres of light of the images. 
\item The confidence intervals are in the range of 10-40\% of the estimated lens parameter values and are increasing to 50-100\% when the uncertainties in the positions of the resolved features exceed one pixel precision.
\end{itemize}

On the whole, we simulated multiple images with realistic extensions and distance scales that have been observed. Taking into account that multiple images with clearly resolved features seem to be more extended than unresolved multiple images, we conclude that the linear transformation matrix between pairs of these images may not be sufficient to yield accurate and precise information and the approach has to be extended including higher order moments in order to become a reliable tool for the model-independent characterisation of gravitational lenses.

\begin{acknowledgements}
JW would like to thank Mauricio Carrasco, Sven Meyer, Robert Reischke, Bj\"{o}rn Malte Sch\"{a}fer, Sebastian Stapelberg, and R\"{u}diger Vaas for helpful discussions and gratefully acknowledges the support by the Deutsche Forschungsgemeinschaft (DFG) WA3547/1-1. NT acknowledges support from the European Research Council in the form of a Consolidator Grant with number 681431.
\end{acknowledgements}
\bibliographystyle{aa}
\bibliography{aa}

\begin{thebibliography}{15}
\expandafter\ifx\csname natexlab\endcsname\relax\def\natexlab#1{#1}\fi

\bibitem[{{Bartelmann}(1996)}]{bib:Bartelmann}
{Bartelmann}, M. 1996, \aap, 313, 697

\bibitem[{{Bartelmann} \& {Schneider}(2001)}]{bib:Weak_lensing_review}
{Bartelmann}, M. \& {Schneider}, P. 2001, \physrep, 340, 291

\bibitem[{{Colley} {et~al.}(1996){Colley}, {Tyson}, \& {Turner}}]{bib:Colley}
{Colley}, W.~N., {Tyson}, J.~A., \& {Turner}, E.~L. 1996, \apjl, 461, L83

\bibitem[{{Donnarumma} {et~al.}(2011){Donnarumma}, {Ettori}, {Meneghetti},
  {Gavazzi}, {Fort}, {Moscardini}, {Romano}, {Fu}, {Giordano}, {Radovich},
  {Maoli}, {Scaramella}, \& {Richard}}]{bib:Donnarumma}
{Donnarumma}, A., {Ettori}, S., {Meneghetti}, M., {et~al.} 2011, \aap, 528, A73

\bibitem[{{Gorenstein} {et~al.}(1984){Gorenstein}, {Shapiro}, {Rogers},
  {Cohen}, {Corey}, {Porcas}, {Falco}, {Bonometti}, {Preston}, {Rius}, \&
  {Whitney}}]{bib:Gorenstein}
{Gorenstein}, M.~V., {Shapiro}, I.~I., {Rogers}, A.~E.~E., {et~al.} 1984, \apj,
  287, 538

\bibitem[{{Kormann} {et~al.}(1994){Kormann}, {Schneider}, \&
  {Bartelmann}}]{bib:Kormann}
{Kormann}, R., {Schneider}, P., \& {Bartelmann}, M. 1994, \aap, 284, 285

\bibitem[{{Meneghetti} {et~al.}(2016){Meneghetti}, {Natarajan}, {Coe},
  {Contini}, {De Lucia}, {Giocoli}, {Acebron}, {Borgani}, {Bradac}, {Diego},
  {Hoag}, {Ishigaki}, {Johnson}, {Jullo}, {Kawamata}, {Lam}, {Limousin},
  {Liesenborgs}, {Oguri}, {Sebesta}, {Sharon}, {Williams}, \&
  {Zitrin}}]{bib:Meneghetti}
{Meneghetti}, M., {Natarajan}, P., {Coe}, D., {et~al.} 2016, ArXiv e-prints
  [\eprint[arXiv]{1606.04548}]

\bibitem[{{Merten} {et~al.}(2015){Merten}, {Meneghetti}, {Postman}, {Umetsu},
  {Zitrin}, {Medezinski}, {Nonino}, {Koekemoer}, {Melchior}, {Gruen},
  {Moustakas}, {Bartelmann}, {Host}, {Donahue}, {Coe}, {Molino}, {Jouvel},
  {Monna}, {Seitz}, {Czakon}, {Lemze}, {Sayers}, {Balestra}, {Rosati},
  {Ben{\'{\i}}tez}, {Biviano}, {Bouwens}, {Bradley}, {Broadhurst}, {Carrasco},
  {Ford}, {Grillo}, {Infante}, {Kelson}, {Lahav}, {Massey}, {Moustakas},
  {Rasia}, {Rhodes}, {Vega}, \& {Zheng}}]{bib:Merten}
{Merten}, J., {Meneghetti}, M., {Postman}, M., {et~al.} 2015, \apj, 806, 4

\bibitem[{{Narayan} \& {Bartelmann}(1996)}]{bib:Narayan}
{Narayan}, R. \& {Bartelmann}, M. 1996, ArXiv Astrophysics e-prints
  [\eprint{astro-ph/9606001}]

\bibitem[{Schneider {et~al.}(1992)Schneider, Ehlers, \& Falco}]{bib:SEF}
Schneider, P., Ehlers, J., \& Falco, E.~E. 1992, {Gravitational Lenses},
  {Astronomy and Astrophysics Library} (New York: Springer)

\bibitem[{{Sharon} {et~al.}(2012){Sharon}, {Gladders}, {Rigby}, {Wuyts},
  {Koester}, {Bayliss}, \& {Barrientos}}]{bib:Sharon}
{Sharon}, K., {Gladders}, M.~D., {Rigby}, J.~R., {et~al.} 2012, \apj, 746, 161

\bibitem[{{Tessore}(2017)}]{bib:Tessore}
{Tessore}, N. 2017, \aap, 597, L1

\bibitem[{{Tessore} {et~al.}(in preparation){Tessore}, {Wagner}, \&
  {Liesenborgs}}]{bib:Tessore2}
{Tessore}, N., {Wagner}, J., \& {Liesenborgs}, J. in preparation

\bibitem[{{Wagner}(2016)}]{bib:Wagner2}
{Wagner}, J. 2016, ArXiv e-prints [\eprint[arXiv]{1612.01793}]

\bibitem[{{Wagner} \& {Bartelmann}(2016)}]{bib:Wagner}
{Wagner}, J. \& {Bartelmann}, M. 2016, Astronomy \& Astrophysics, 590, A34

\end{thebibliography}

%%%%%%%%%%%%%%%%%%%%%%%%%%%%%%%%%%%%%%%%%%%%%%%%%%%%%%%%%%%%%%%%%%%%%%%%%%%%%
\appendix
\section{Solution to the system of Equations~\ref{eq:T11_phi} to \ref{eq:T22_phi}}
\label{app:phi_solution}

Given an observation of three images $A, B, C$, with $A$ being the reference image, in the coordinate system described in Section~\ref{sec:equivalence_of_approaches}, the solution of the system of Equations~\ref{eq:T11_phi} to \ref{eq:T22_phi} is calculated as
\begin{align}
\tilde{\phi}_{12}^{(A)} &= \dfrac{T_{12}^{(AB)}T_{21}^{(A,C)} - T_{21}^{(A,B)}T_{12}^{(A,C)}}{\left(T_{11}^{(A,B)} - T_{22}^{(A,B)} \right) T_{21}^{(A,C)} -\left(T_{11}^{(A,C)} - T_{22}^{(A,C)} \right) T_{21}^{(A,B)} } \;, \label{eq:sol1} \\
\tilde{\phi}_{22}^{(A)} &= \dfrac{\left(T_{11}^{(A,B)} - T_{22}^{(A,B)} \right) T_{12}^{(A,C)} -\left(T_{11}^{(A,C)} - T_{22}^{(A,C)} \right) T_{12}^{(A,B)} }{\left(T_{11}^{(A,B)} - T_{22}^{(A,B)} \right) T_{21}^{(A,C)} -\left(T_{11}^{(A,C)} - T_{22}^{(A,C)} \right) T_{21}^{(A,B)} } \;, \label{eq:sol2} \\
f_\phi^{(I)} &= \dfrac{\det\left( T^{(A,I)}\right)}{T_{22}^{(A,I)} -\tilde{\phi}_{12}^{(A)} T_{21}^{(A,I)} }\;,  \quad I = B, C \;, \label{eq:sol3} \\ 
\tilde{\phi}_{12}^{(I)} &= \dfrac{T_{11}^{(A,I)}\tilde{\phi}_{12}^{(A)} -T_{12}^{(A,I)} }{T_{22}^{(A,I)} -\tilde{\phi}_{12}^{(A)} T_{21}^{(A,I)}} \;, \quad I = B, C\;, \label{eq:sol4} \\
\tilde{\phi}_{22}^{(I)} &= \dfrac{T_{11}^{(A,I)}\tilde{\phi}_{22}^{(A)} -T_{12}^{(A,I)}\tilde{\phi}_{12}^{(A)} }{T_{22}^{(A,I)} -\tilde{\phi}_{12}^{(A)} T_{21}^{(A,I)}} \;,\quad I = B, C \;. \label{eq:sol5}
\end{align}
To arrive at these equations, we use
\begin{equation}
\tilde{\phi}_{12}^{(A)} \left( T_{11}^{(A,I)} - T_{22}^{(A,I)} \right) + \tilde{\phi}_{22}^{(A)} T_{21}^{(A,I)} = T_{12}^{(A,I)} \;, \quad I =B, C
\label{eq:fixed_curl}
\end{equation}
as done analogously for the ratios of convergences and reduced shear components in \cite{bib:Tessore}.

\end{document}